\documentclass[12pt]{article}
\usepackage{a4wide,epsfig,amsmath,amssymb,cite,scalefnt}
\usepackage{feynarts}

\newcommand{\lnMgl}{L_{\tilde g}}
\newcommand{\lnMsq}{L_{\tilde{b}_i}}

\newcommand{\lnMt}{L_t}

\newcommand{\Mgl}{m_{\tilde g}}
\newcommand{\Msq}{m_{\tilde{b}_i}}

\newcommand{\mugut}{\mu_{\abbrev\rm GUT}}

\newcommand{\abbrev}{\scalefont{.9}}
\newcommand{\drbar}{$\overline{\mbox{\abbrev DR}}$}
\newcommand{\msbar}{$\overline{\mbox{\abbrev MS}}$}
\newcommand{\drbarmath}{\overline{\rm\abbrev DR}}

\newcommand{\asDR}[1]{\alpha_s^{\drbarmath{}#1}}

\newcommand{\lt}{L_{t}}
\newcommand{\lgl}{L_{\tilde{g}}}
\newcommand{\lsq}{L_{\tilde{q}}}

\newcommand{\mt}{m_t}
\newcommand{\mg}{m_{\tilde{g}}}
\newcommand{\msq}{m_{\tilde{q}}}

\begin{document}

%%%%%%%%%%%%%%%%%%%%%%%%%%%%%%%%%%%%%%%%%%%%%%%%%%%%%%%%%%%%%%%%%%%%%%%%%%%%

%- }}}
%- {{{ title and abstract

\title{\vskip-3cm{\baselineskip14pt
    \begin{flushleft}
      \normalsize SFB/CPP-08-75 \\
      \normalsize TTP/08-25  
  \end{flushleft}}
  \vskip1.5cm
  Matching coefficients  for $\alpha_s$ and $m_b$ to ${\cal O}(\alpha_s^2)$ \\
  in the MSSM }
\author{ A. Bauer,
  L. Mihaila, J. Salomon\\
  {\small\it Institut f{\"u}r Theoretische Teilchenphysik,
    Universit{\"a}t Karlsruhe (TH),}\\
{\small\it Karlsruhe Institute of Technology (KIT)}\\
  {\small\it 76128 Karlsruhe, Germany}\\
}

\date{}

\maketitle

\thispagestyle{empty}

\begin{abstract}
We compute the exact two-loop matching coefficients for the strong
coupling constant 
$\alpha_s$ and the bottom-quark mass $m_b$ within the Minimal
Supersymmetric Standard Model (MSSM), taking into account
 ${\cal  O}(\alpha_s^2)$  
contributions  from Supersymmetric Quantum Chromodynamics (SQCD).
We find that the  
explicit mass pattern  of the   supersymmetric
particles has a  significant impact on the
predictions of $\alpha_s$ and $m_b$ at high energies. Further on, the
three-loop 
corrections  exceed the uncertainty
due to the current experimental accuracy.  In case of the
 the running  bottom-quark mass, they can reach in the  large $\tan
 \beta$ regime  up to $30\%$ of the tree-level value.
\medskip

\noindent
PACS numbers: 11.30.Pb, 12.38.-t, 12.38.Bx, 12.10.Kt

\end{abstract}

%- }}}
%- {{{ Introduction:

\section{Introduction}

Supersymmetry (SUSY) is currently believed to play an important role in
physics beyond the Standard Model (SM). 
 A compelling argument in
 favour of SUSY is the particle content of the
 MSSM, that leads in a natural way to the unification of the three gauge
 couplings at a high energy scale $\mu\simeq
10^{16}$~GeV, in agreement with predictions of Grand Unification Theories 
(GUT)~\cite{Dimopoulos:1981yj,Ibanez:1981yh,Amaldi:1991cn}.
\\
Apart from the gauge coupling
unification, in GUT models based on simple groups such as
$SU(5)$\cite{Georgi:1974sy} or $SO(10)$~\cite{Fritzsch:1974nn}, also
 the  bottom ($y_b$) and  tau ($y_{\tau}$) Yukawa
 couplings  unify  at 
the GUT scale. For some models based on  $SO(10)$ (or larger groups)
even the unification of the  bottom, tau and top ($y_t$) Yukawa
 couplings is predicted. However, the condition of Yukawa coupling  unification
can be fulfilled within the MSSM only for  two regions  of  $\tan \beta$,
the ratio of Higgs field  vacuum expectation values: $\tan \beta \approx
1$ and $\tan \beta \approx
50$~\cite{Langacker:1994bc, Hall:1993gn, DiazCruz:2000mn}.  A main feature of SUSY models 
with large $\tan \beta$ is that the supersymmetric radiative corrections
to fermion masses and couplings can be as large as  the leading order
(LO) contributions ~\cite{Hall:1993gn, Hempfling:1993kv,Carena:1994bv}. This
renders the knowledge of the higher order (HO) corrections in
perturbation theory  
mandatory. On the other hand, the unification condition
becomes very sensitive to  the low energy
parameters~\cite{Tobe:2003bc}.  This property can be
exploited to greatly constrain the  allowed MSSM parameter space.

With the advent of the CERN
Large Hadron Collider (LHC), we will be able to probe the realization of
SUSY in nature  to energy scales of ${\cal O}(1)$~TeV. 
 In particular, precision measurements and computations will allow to test  the
 low-energy supersymmetric relations between the particle masses and
 couplings. It is often argued (for reviews see
e.g.\ Refs.~\cite{Blair:2002pg,Aguilar-Saavedra:2005pw}) that, from the
precise knowledge of the  low-energy supersymmetric parameters one 
 can shed light on the origin and
mechanism of supersymmetry breaking  and even on physics at much higher
energies, like the GUT scale. The
extrapolation of the supersymmetric parameters measured at the TeV energy
scale to  the GUT-scale raises inevitably the question of
uncertainties  involved. Currently, there are  four publicly
available  spectrum generating codes~\cite{Paige:2003mg,
  Allanach:2001kg, Porod:2003um, Djouadi:2002ze}  based on two-loop order MSSM
renormalization group equations (RGEs)~\cite{Martin:1993yx,
  Martin:1993zk, Jack:1994kd, Yamada:1994id} subjected to two types of
 boundary conditions. One set of constraints accounts for the weak-scale
matching between the MSSM and SM parameters to one-loop
order~\cite{Pierce:1996zz}. The second 
one allows for the SUSY breaking at the high scale according to   
specific models like minimal supergravity, gauge mediation and 
  anomaly mediation. The approximations within the codes differ
by higher order corrections and by the treatment of the low-energy
threshold corrections. The typical spread of the results is within
few percents, which does not always meet the experimental
accuracy~\cite{Allanach:2003jw}. Along the same line,   recent
analyses~\cite{Harlander:2005wm, Harlander:2007wh} have proven that 
the three-loop order effects on the running of the strong
coupling constant $\alpha_s$ and the bottom quark
mass $m_b$
may  exceed those induced by the  
 current experimental accuracy~\cite{Bethke:2006ac, Kuhn:2007vp}. \\
Very recently, Ref.~\cite{Noth:2008tw} computed the two-loop SQCD and
top-induced  Supersymmetric Electroweak (SEW) 
corrections to the effective bottom Yukawa couplings. The knowledge of
the two-loop corrections
allows predictions for the branching ratios of the MSSM Higgs bosons
with  per-cent level accuracy.

The aim of this paper 
is to compute  the weak-scale   
matching relations for the strong coupling constant and the bottom
quark mass with two-loop accuracy, taking into account the exact dependence on the particle
masses. This will extend the study of 
Ref.~\cite{Harlander:2005wm} allowing
phenomenological analyses based on   quasi-realistic  mass 
spectra for SUSY particles, with  three-loop order accuracy. 
 However, we will consider in this paper only the ${\cal
   O}(\alpha_s^2)$ corrections from 
 SQCD and  postpone the study of the SEW contributions ${\cal
   O}(\alpha_s y_t^2, \alpha_s y_b^2,  y_t^4, y_t^2 y_b^2,  y_b^4 )$
     to further investigations. Whereas the  SEW  corrections to the
     decoupling coefficient of $\alpha_s$ are 
 expected to be negligible, their numerical impact on the running bottom
 quark mass can become as important as those from SQCD.
At the one-loop order~\cite{Pierce:1996zz}, the main contributions to the
 bottom-quark mass 
 decoupling coefficient arise from diagrams
 containing gluino- and Higgsino-exchange. In the most regions of the
 MSSM parameter space the gluino contribution is the dominant one and
  can be as large as the tree-level bottom quark mass.
  However,
there are  domains in the parameter space where the corrections due to
gluino- and Higgsino-exchange can become of the same  order and have 
opposite sign. These regions contain the special  points for which the
Yukawa coupling  unification occurs ~\cite{Tobe:2003bc, Blazek:2002ta}.
For the MSSM parameters for which the radiative corrections to the
bottom quark mass are comparable with the LO ones,
Ref.~\cite{Carena:1999py} proposed a method to  
 resum them to all orders in perturbation theory.  
A numerical comparison with
the results of Ref.~\cite{Carena:1999py} can be found in
Section~\ref{sec::num}.
 
The paper is organized as follows: in Section~\ref{sec::frame} and
Section~\ref{sec::ren} we discuss the theoretical framework and the
renormalization scheme we use. In   Section~\ref{sec::res} we present the
analytical one- and two-loop results. The latter ones are displayed in
analytical form  
for  three simplifying mass hierarchies among the SUSY particles. The
numerical effects are studied in Section~\ref{sec::num}.

%- }}}

%- {{{framework:
\section{\label{sec::frame}Framework}
As already stated above, the underlying motivation for the running analysis is
to relate physical parameters measured at the electroweak scale with the
Lagrange parameters at the GUT scale. The running parameters are most
conveniently defined in mass-independent renormalization schemes such as
\msbar{}~\cite{Bardeen:1978yd} for the SM parameters and
\drbar{}~\cite{Siegel:1979wq} for  
the MSSM parameters. It is 
well known that in such  ``unphysical''  renormalization schemes the
Appelquist-Carazzone decoupling theorem~\cite{Appelquist:1974tg}  does
not hold in its naive form. Quantum corrections  
to low-energy processes contain logarithmically enhanced
contributions from heavy particles with masses  much greater than the
energy-scale  of the process under consideration. An elegant approach to
get rid of this unwanted behaviour in the \msbar{} or \drbar{} scheme  is to
formulate an effective 
theory (ET) ~\cite{Chetyrkin:1997un, Steinhauser:2002rq} integrating out
all heavy particles.
 The coupling constants  defined within the ET  must be  modified in order to
account for the effects of the heavy fields. They are related to the
parameters  of the full theory by  the so-called matching or decoupling
relations.

For moderate mass splittings between SUSY particles, {\it i.e.}\ there
are no large   
logarithms in the theory that have to be resummed by means of RGEs,  the
decoupling of heavy particles might be performed in one
step~\cite{Ferreira:1996ug}.  The 
 energy-scale at which the decoupling is performed is not fixed by the
 theory. It is usually chosen to be $\mu \simeq \tilde M $, where
 $\tilde M$ is a typical 
 heavy particle mass. 
The MSSM parameters at energies $E\simeq \tilde M$ can be determined
from the knowledge of the corresponding 
SM parameters and the associated  decoupling relations. 

 The decoupling coefficients for the strong coupling constant and for
 the light  quark masses are known in QCD with
 four-~\cite{Schroder:2005hy,Chetyrkin:2005ia} and
 three-loop~\cite{Chetyrkin:1997un} accuracy, 
 respectively.
 Due to the 
 presence of many  mass scales, their computation within SQCD and SEW
 is quite involved.  At one-loop order, they are known exactly
 ~\cite{Pierce:1996zz,Bednyakov:2007vm}. At  two-loop order, the
 decoupling coefficient 
 for the strong coupling is known only for specific mass
 hierarchies~\cite{Harlander:2005wm}. Recently, the two-loop SQCD
 corrections for the
 decoupling coefficient of the bottom-quark mass was
 computed~\cite{Bednyakov:2007vm}. 
 The focus of this paper is the analytical  computation of the decoupling
relations for the  strong coupling constant  and  the bottom-quark mass
within SQCD through two-loops using a different method as
Ref.~\cite{Bednyakov:2007vm}. The comparison of the results will be
discussed in the next Section.\\

\begin{figure}[t]
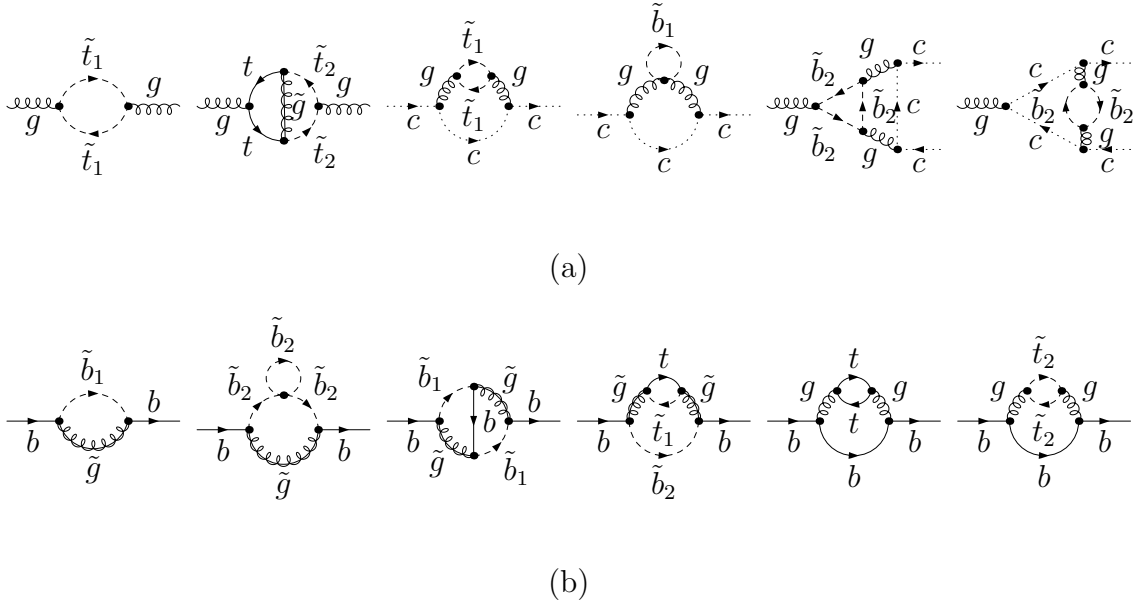

  \begin{center}
    \begin{tabular}{c}
\unitlength=1.bp%
\begin{feynartspicture}(430,100)(6,1)
\input  figs/diagcc
\end{feynartspicture}\\
(a)\\
\unitlength=1.bp%
\begin{feynartspicture}(430,100)(6,1)
\input  figs/diaqq
\end{feynartspicture}\\
(b)
    \end{tabular}
    \parbox{14.cm}{
 \caption[]{\label{fig::diagrams}\sloppy Sample diagrams
        contributing to $\zeta_3$, $\tilde{\zeta}_3$, $\tilde{\zeta}_1$
        and $\zeta_m$ with 
        gluons ($g$), ghosts ($c$),  bottom/top quarks ($b/t$), bottom/top squarks
        ($\tilde{b}/\tilde{t}$) and gluinos ($\tilde{g}$).    }
}
  \end{center}
\end{figure}

We consider SQCD with $n_f$ active quark and $n_s=n_f$ active squark flavours
and $n_{\tilde g}=1$ gluinos. Furthermore, we assume that $n_l=5$ quarks
are light (among which the bottom quark) and that the top quark and
all squarks and the gluino are heavy.
Integrating out the heavy fields from the full SQCD Lagrangian, we
obtain 
the Lagrange density corresponding to the effective QCD with $n_l$
light quarks plus non-renormalizable interactions. The latter are
suppressed by negative powers   of the heavy masses and  will be
neglected here. The effective Lagrangian can be written as follows:

\begin{eqnarray}
{\cal L^\prime}(g_s^{0}, m_q^{0}, \xi^{0}; \psi_q^{0},G_{\mu}^{0,a},
c^{0,a}; \zeta_i^{0}) = {\cal L}^{SQCD}(g_s^{0 \prime}, m_q^{0\prime},
\xi^{0\prime}; \psi_q^{0\prime},G_{\mu}^{0\prime,a},c^{0\prime,a})\,, 
 \label{eq::lag}
\end{eqnarray} 
where $ \psi_q, G_{\mu}^a ,c^a $ denote the light-quark, the gluon and
the ghost 
fields, respectively, $m_q$ stands for the light quark masses, $\xi$ is
the gauge parameter and $g_s=\sqrt{4\pi \alpha_s}$ is the strong
coupling. The index $0$ 
marks bare quantities.
${\cal L}^{SQCD}$ is the usual SQCD Lagrangian from which all heavy
fields have been discarded. As a result the fields, masses and
couplings  associated with light particles have to be rescaled.  They
are  labeled by a prime in Eq.~(\ref{eq::lag}) and are related with the
original parameters through  decoupling relations:
\begin{eqnarray}
  g_s^{0\prime}  =\zeta_g^0 g_s^0   \,,&\quad
  m_q^{0\prime}  =\zeta_m^0m_q^0    \,,&\quad
  \xi^{0\prime}-1=\zeta_3^0(\xi^0-1)\,,
  \nonumber\\
  \psi_q^{0\prime} =\sqrt{\zeta_2^0}\psi_q^0     \,,&\quad
  G_\mu^{0\prime,a}=\sqrt{\zeta_3^0}G_\mu^{0,a}  \,,&\quad
  c^{0\prime,a}    =\sqrt{\tilde\zeta_3^0}c^{0,a}\,.
  \label{eq::bare_dec}
\end{eqnarray}
Refs.~\cite{Chetyrkin:1997un}  showed that the  bare decoupling constants 
$\zeta_m^0,\, \zeta_2^0,\, \zeta_3^0, 
\tilde\zeta_3^0 $ can be  derived from the quark, the gluon and the ghost
propagators, all evaluated at vanishing external momenta. As a result,
for calculations performed within the framework of Dimensional
Regularization/Reduction (DREG/DRED)
only the diagrams containing at least one  heavy particle inside
the loops do not vanish. In  Fig.~\ref{fig::diagrams} are shown 
 sample Feynman diagrams contributing to the
decoupling coefficients for the strong coupling (a) and the bottom-quark
mass~(b). 

For the
computation of $\zeta_g$ one has to use the well-known Ward
identities. A convenient choice is the relation:
\begin{eqnarray}
  \zeta_g^0 &=& \frac{\tilde{\zeta}_1^0}{\tilde{\zeta}_3^0\sqrt{\zeta_3^0}}
  \,,
  \label{eq::zetag0}
\end{eqnarray}
where $\tilde{\zeta}_1^0$ denotes the decoupling constant for the
ghost-gluon vertex.\\
The finite decoupling coefficients are obtained upon the
renormalization of the bare parameters. They are
given by
\begin{eqnarray}
  \zeta_g = \frac{Z_g}{Z_g^\prime} \zeta_g^0
  \,,\quad  \zeta_m = \frac{Z_m}{Z_m^\prime} \zeta_m^0\,,
  \label{eq::zetagren}
\end{eqnarray}
where $Z_g^\prime$ and $Z_m^\prime$ correspond to the renormalization
constants in the 
effective theory. Since we are interested in the two-loop results for
$\zeta_i,\, i=g,m$, the corresponding renormalization constants for
SQCD and QCD have to be implemented with the same accuracy. Analytical 
results for them can be found in
Refs.~\cite{ Jack:1994kd, Steinhauser:2002rq, Bednyakov:2002sf}.

As mentioned above, the
decoupling coefficients can be related with vacuum integrals. 
The latter can be recursively reduced to a
master-integral~\cite{Davydychev:1992mt} using the method of integration
by parts~\cite{Chetyrkin:1981qh}. Given the large number of diagrams and 
occurrence 
of many different mass scales, we  computed them with the help of an
automated setup. The Feynman diagrams were generated with
{\tt QGRAF}~\cite{Nogueira:1991ex} 
and further processed with   {\tt q2e}~\cite{Seidensticker:1999bb}. The
reduction  of various 
vacuum integrals to the master
integral was performed by a self written {\tt
  FORM}~\cite{Vermaseren:2000nd} routine. The reduction of  
topologies with two different massive  and one massless lines
requires a careful treatment.  The related master integral
can be easily derived from its general expression  valid for massive
lines, given in
Ref.~\cite{Davydychev:1992mt}.

\section{\label{sec::ren} Regularization and renormalization scheme}

In our setup, we used the squark mass
eigenstates and their  mixing angles as input parameters.
For convenience, we give below the relations between  them  and the parameters  of the MSSM
Lagrangian.\\
 The squark mass eigenstates $\tilde{q}_{1,2}$ and their mass
 eigenvalues $m_{\tilde{q}_{1,2}}$  are obtained by 
diagonalizing the  mass matrix
\begin{eqnarray}
{\cal M}_{\tilde q}= 
\left(
\begin{array}{cc}
M_L^2& m_q X_q\\
m_q X_q& M_R^2
\end{array}
\right)\,,
\label{eq::mixing}
\end{eqnarray}
where we used the notation
\begin{eqnarray}
 X_q &=& A_q-\mu \left\{\begin{array}{ll}
\tan\beta\,, &\mbox{ for  down-type quarks}\\
\cot\beta\,,  &\mbox{ for  up-type quarks}
\end{array}\right . \,,\nonumber\\
M_L^2 &=& M_{\tilde Q}^2+m_q^2+M_Z^2 (I_3^q-Q_q s_W^2)\cos
2\beta\,,\nonumber\\ 
M_R^2 &=& M_{\tilde{D},\tilde{U}}^2+m_q^2+M_Z^2 Q_q s_W^2\cos 2\beta\,.
\end{eqnarray}
Here $m_q$, $I_3^q$ and $Q_q$ are the mass,  isospin and electric charge
of the quark $q$, respectively, and $s_W=\sin\theta_W$. The parameters
$M_{\tilde Q}$ and $M_{\tilde{D},\tilde{U}}$ are the soft supersymmetry breaking masses,
$A_q$ is  a trilinear coupling and $\mu $   is the
Higgs-Higgsino bilinear coupling. 

The squark mass eigenvalues  are defined through the unitary transformation
\begin{eqnarray}
\left(
\begin{array}{cc}
m_{\tilde{q}_1}^2& 0\\
0& m_{\tilde{q}_2}^2
\end{array}
\right) = {\cal R}_{\tilde q} {\cal M}_{\tilde q} {\cal R}_{\tilde
  q}^\dag\,,\quad \mbox{with} \quad{\cal R}_{\tilde q} =\left(
\begin{array}{cc}
\cos \theta_q & \sin \theta_q \\
-\sin \theta_q& \cos \theta_q
\end{array}
\right)\,,
\end{eqnarray}
and the squark mixing angle  through 
\begin{eqnarray}
\sin 2\theta_q =  \frac{2 m_q X_q}{m_{\tilde{q}_1}^2-m_{\tilde{q}_2}^2} \,.
\label{eq::mixb} 
\end{eqnarray}

Since we consider the two-loop ${\cal O}(\alpha_s^2)$ corrections, only the
one-loop  ${\cal O}(\alpha_s)$ counter\-terms for the input parameters are
required.  We have
chosen the \drbar{} scheme to renormalize the strong coupling constant, and
the   on-shell scheme for the
masses of the heavy particles, {\it i.e.\ } the gluino,  squarks and  top
quark. The corresponding one-loop renormalization constants  are  known
analytically  (see, {\it e.g.\ }, Ref.~\cite{Pierce:1996zz}). 
For the computation of the two-loop $\zeta_{m_b}$,  also the one-loop
counter\-term for the sbottom
mixing angle $\theta_b$   is required.
 We
adopted the on-shell renormalization prescription as defined in
Ref.~\cite{Guasch:1998as}
\begin{eqnarray}
\delta\theta_{b}&=&\frac{\mbox{Re}
  \Sigma_{\tilde{b}_{12}}(m^2_{\tilde{b}_1})+\mbox{Re}
  \Sigma_{\tilde{b}_{12}}(m^2_{\tilde{b}_2})}{2
  (m^2_{\tilde{b}_1}-m^2_{\tilde{b}_2})}\,, 
\end{eqnarray}
 where $\Sigma_{\tilde{b}_{12}}$ is the non-diagonal on-shell sbottom
 self-energy. 

As we neglect the bottom-quark mass w.r.t.\ heavy particle masses, an
explicit dependence of the radiative corrections on $m_b$ can occur only
through bottom Yukawa couplings. In order to avoid the occurrence of
large logarithms of the form 
$\alpha_s^2\log(\mu^2/m_b^2)$ with $\mu\simeq \tilde M$, we have renormalized
the bottom quark mass in the \drbar{} scheme. In this way, the large
logarithms are absorbed into the running mass and the higher
order corrections are maintained small.

The renormalization prescription  for the trilinear coupling $A_b$ is fixed
by the tree-level relation Eq.~(\ref{eq::mixb}), while the parameters
$\mu$ and $\tan\beta$ do not acquire 
${\cal O}(\alpha_s)$ corrections to the one-loop level.
Generically, the counterterm for $A_b$ can be expressed as
\begin{eqnarray}
\delta A_b = \left(2 \cos 2\theta_b
  \delta\theta_{b}+\sin 2\theta_b \frac{ \delta m^2_{\tilde{b}_1}-\delta
    m^2_{\tilde{b}_2}}{m^2_{\tilde{b}_1}-m^2_{\tilde{b}_2}}-\sin
  2\theta_b\frac{\delta m_b}{m_b}\right)
\frac{m^2_{\tilde{b}_1}-m^2_{\tilde{b}_2}}{2
  m_b} \,,
\label{eq::dab}
\end{eqnarray}
where $\delta m_b$ and $ \delta m^2_{\tilde{b}_{1,2}} $ 
are the counterterms corresponding to bottom-quark and squark masses, respectively.
Due to the use of different renormalization prescriptions for the
bottom/sbottom masses and mixing angle, the parameter $A_b$ is renormalized
in a {\it mixed} scheme. 
%For the numerical analysis, we follow the
%method of Ref.~\cite{Heinemeyer:2004xw} to 
%determine the  one-loop relations between the values of the parameter
%$A_b$ defined in different renormalization schemes. 
\\

For the regularization of ultra-violet divergencies, we have implemented
DRED with the help of the so-called
$\epsilon$-scalars~\cite{Jack:1993ws}. 
In softly broken SUSY theories, as it is the case of MSSM/SQCD, they get
a radiatively induced mass. We choose to renormalize their mass in the
on-shell scheme, requiring that the renormalized mass is equal to 
zero. \\
There are also other approaches available in the literature. We want to
mention the one proposed in Ref.~\cite{Jack:1994rk}, where
the $\epsilon$-scalars are treated as massive particles. This
approach is known in the literature as the \drbar{}$^\prime$ scheme. 
 In this case,
the $\epsilon$-scalars have to be decoupled together with the  heavy
particles of the theory~\cite{Bednyakov:2007vm}. The
advantage of this method is  that it directly 
relates SQCD parameters regularized within DRED with those of QCD
regularized within DREG, which are known from experiments. The  price of
this ``shortcut'' is on one hand, that   
 additional diagrams containing the $\epsilon$-scalars as
massive particles occur in the calculation of the decoupling
coefficients. On the other hand, the contributions originating from
the change of regularization scheme and those from the decoupling of
heavy particles  are not distinguishable anymore.

 In our approach the change of regularization scheme from DRED to DREG
has to be performed explicitly within  QCD~\cite{Martin:1993zk,
  Jack:1994kd, Harlander:2006rj}. For our
purposes, the two-loop conversion relations for the strong coupling
constant and the bottom-quark mass are required. The complication that
arises at this stage is the occurrence of the {\it evanescent coupling}
$\alpha_e$ associated with  the  $\epsilon$-scalar-quark-quark vertex. 
This has to be distinguished from the gauge coupling within
non-supersymmetric theories (e.g.\ QCD). However, in SQCD they  are
identical with 
the  gauge couplings, as required by SUSY. Using the ET procedure, we
can relate them with the strong coupling within the full theory with the
help of 
decoupling  relations similar with those introduced in
Eq.~(\ref{eq::bare_dec})

\begin{eqnarray}
 \alpha_e^{\prime}=\zeta_e \alpha_e= \zeta_e \alpha_s\,.
  \label{eq::zetae}
\end{eqnarray}
Following the method described above, one can calculate $\zeta_e$
evaluating the $\epsilon$-scalar and quark-propagators and the
$\epsilon$-scalar-quark-quark vertex at vanishing external momentum. In
Ref.~\cite{Harlander:2007wh}, its  one-loop expression was computed under the
simplifying assumption of a degenerate SUSY-mass spectrum. In principle,
for our numerical analyses, that rely on solving  a system of coupled
differential equations involving also the evanescent coupling $\alpha_e$,
even the  two-loop order corrections  to $\zeta_e $ are needed. However,
from the explicit calculation it turned out that the numerical effects
induced by the    two-loop corrections  to $\zeta_e$ are  below the
per-mille level. For simplicity, we do not display  the 
corresponding   two-loop results in the following. The analytical formulae   are available
upon request from the authors.

The method outlined here and the one introduced in
Ref.~\cite{Bednyakov:2007vm} for the calculation of the decoupling
coefficient of the bottom-quark
mass are 
equivalent. This has to be understood in the usual sense, that the
predictions  for physical observables
made in one scheme at a given order in perturbation theory can be translated to the other scheme through
redefinitions of  masses and couplings. We have explicitly checked
 implementing additionally the method of Ref.~\cite{Bednyakov:2007vm} in our setup the
equivalence  property for the  decoupling  coefficient of the bottom-quark mass
$\zeta_m$ through two-loop order. Apart from  the obvious rescaling of the
strong coupling and the bottom quark mass, also the sbottom masses have 
to be modified ~\cite{Jack:1994rk}
\begin{eqnarray}
m_{\tilde{b}}^2|_{\drbarmath{}^\prime}=m_{\tilde{b}}^2|_{\drbarmath{}}
-\frac{\asDR{}}{2\pi} C_F m_{\epsilon}^2\,.
\end{eqnarray}
Here $C_F$ is the Casimir invariant in the fundamental representation
and $m_{\epsilon}$ denotes the mass of the  $\epsilon$-scalars. The
indices \drbar{} and \drbar{}$^\prime$, respectively, specify the
regularization scheme. We also compared numerically the results for the
two-loop $\zeta_{m_b}$ obtained using  our method with the ones depicted in
Fig.~3 of Ref.~\cite{Bednyakov:2007vm} and found very good agreement.

\section{\label{sec::res} Analytical results}
\subsection{\label{sec::oneloop}One-loop results}

The exact one-loop results for the decoupling coefficients of the strong
coupling constant $\zeta_s$ 
and  bottom-quark mass $\zeta_m$ can be found in
Refs.~\cite{Pierce:1996zz, Bednyakov:2007vm}. The analytical formula for
$\zeta_e$ is new and we give it below up to $\cal{O}(\epsilon)$.
{\allowdisplaybreaks
\begin{align}
%\begin{eqnarray}
\zeta _s =&
%& 
1+\frac{\alpha_s^{\rm{ (SQCD)}}}{ \pi }
\Bigg[-\frac{1}{6} C_A L_{\tilde{g}}-\frac{1}{6}L_t
-\sum_{q}\sum_{i=1,2}\frac{1}{24} L_{\tilde{q}_i}
  \nonumber \\ 
%&
&-\epsilon  \left(\frac{C_A}{12}
  \left(L_{\tilde{g}}^2+\zeta(2)\right) 
  +\frac{1}{12} \left(
  L_t^2+\zeta(2)\right)-\frac{1}{48}\sum_{q}\sum_{i=1,2}\left( 
    L_{\tilde{q}_i}^2+ \zeta(2)\right)\right)\Bigg]\,,
\label{eq::zetas1l}
\\
 \zeta_{e, q} =&
%& 
1+ \frac{\alpha_s^{\rm{ (SQCD)}}}{ \pi } \Bigg\{-T_F
 \frac{ L_t}{2}  +\frac{C_A}{4} \Bigg(   
2+L_{\tilde{g}} + \sum_{i=1,2}\left(L_{\tilde{g}}-L_{\tilde{q}_i}\right)
\frac{m_{\tilde{q}_i}^2}{m_{\tilde{g}}^2-m_{\tilde{q}_1}^2}\Bigg)  
\nonumber\\ 
%&
&+
  \frac{C_F}{4} \Bigg( 
\sum_{i=1,2} \left(-1-2 L_{\tilde{g}}+2 L_{\tilde{q}_i}
+\left(-L_{\tilde{g}}+L_{\tilde{q}_i}\right) 
\frac{m_{\tilde{q}_i}^2}{m_{\tilde{g}}^2 - m_{\tilde{q}_i}^2} \right) \frac{
  m_{\tilde{q}_i}^2}{ m_{\tilde{g}}^2 -m_{\tilde{q}_i}^2}
\nonumber\\
%&
&+ \left(-3-2 L_{\tilde{g}}\right) \Bigg)
+ \epsilon \Bigg[-  \frac{T_F}{4} \left(L_t^2+\zeta(2)\right)+\frac{C_A}{8}
  \Bigg( 4+4 L_{\tilde{g}}+L_{\tilde{g}}^2+\zeta(2) +
\nonumber\\ 
%&
&+
\frac{1}{2}\sum_{i=1,2} \left(L_{\tilde{g}}-L_{\tilde{q}_i}\right)
\left(2+L_{\tilde{g}}+L_{\tilde{q}_i}\right)\frac{ m_{\tilde{q}_i}^2}{
  m_{\tilde{g}}^2-m_{\tilde{q}_i}^2} \Bigg)
+
\frac{C_F}{8} \Bigg( -7-6 L_{\tilde{g}}
-2 L_{\tilde{g}}^2
\nonumber\\ 
%&
&
-2 \zeta(2)
+ \frac{1}{2}\sum_{i=1,2} \Bigg(
-3-6 L_{\tilde{g}}-2 L_{\tilde{g}}^2+4 L_{\tilde{q}_i}+2
L_{\tilde{q}_i}^2
\nonumber\\ 
% &
&-
 \left(L_{\tilde{g}}-L_{\tilde{q}_1}\right)
\left(3+L_{\tilde{g}}+L_{\tilde{q}_1}\right)\frac{ m_{\tilde{q}_i}^2}{ 
m_{\tilde{g}}^2-m_{\tilde{q}_i}^2 }\Bigg)\frac{ m_{\tilde{q}_i}^2}{ 
m_{\tilde{g}}^2-m_{\tilde{q}_i}^2 }\Bigg)\bigg] \Bigg\}\,,
\label{eq::zetae1l}
\\
\zeta_{m_b} = &
% & 
 1 + \frac{\alpha_s^{\rm{ (SQCD)}}}{ \pi }
C_F\sum_{i=1,2}\bigg\{-\frac{(1 + \lnMsq)}{4}\frac{ 
  \Msq^2}{ (\Msq^2-\Mgl^2)} 
+  \frac{ (3 +
     2 \lnMsq) \Msq^4-(3 + 2 \lnMgl) \Mgl^4 }{16 (\Msq^2-\Mgl^2)^2}
 \nonumber\\
%&
&-
\frac{(-1)^i\, X_b \Mgl }{m_{\tilde{b}_1}^2-m_{\tilde{b}_2}^2}\frac{ 
\Msq^2  \lnMsq  - \Mgl^2  \lnMgl}{2 (\Msq^2-\Mgl^2)}  + 
  \epsilon \bigg[-\frac{\Msq^2 (2 + \lnMsq (2 + \lnMsq) +\zeta(2)
   )}{8 (\Msq^2-\Mgl^2)} 
\nonumber\\
%&
&+
 \frac{ 
     \Msq^4 (7 + 2 \lnMsq (3 + \lnMsq) + 2 \zeta(2))-\Mgl^4 (7 + 2
     \lnMgl (3 + \lnMgl) + 2 \zeta(2)) }{32 (\Msq^2-\Mgl^2)^2}  
 \nonumber\\
%&
&+ 
   \frac{(-1)^i\, X_b \Mgl}{m_{\tilde{b}_1}^2-m_{\tilde{b}_2}^2}
   \frac{ \Mgl^2  \lnMgl (2 
+ \lnMgl)  -  \Msq^2  \lnMsq (2 + \lnMsq) }{4
     (\Msq^2-\Mgl^2)} \bigg] 
       \bigg\}\,.
\label{eq::zetam1l}
\end{align}
%\end{eqnarray}
}

In Eqs.~(\ref{eq::zetas1l}), (\ref{eq::zetae1l}), and (\ref{eq::zetam1l}), we have
adopted the abbreviations 
\begin{equation}
L_i = \ln\frac{\mu^2}{m_i^2}\,,\quad i \in
\{t,\tilde{g},\tilde{q}_{1,2},\tilde{b}_{1,2}\}\,, 
\label{eq::abbv}
\end{equation} 
 where  $\tilde{q}_{i} (\tilde{b}_{i})$ denote the super-partners of
 the quark $q (b)$.
   For our purposes,  the special case $\zeta_{e, q=b}$ is of
 interest. 
\\
 The 
 colour factors are defined in case of a gauge group $SU(N)$ as follows  
\begin{equation}
C_F = \frac{N^2-1}{2 N}\,\quad C_A=N\,, \quad T_F=\frac{1}{2}\,.
\end{equation} 
 Furthermore, we used the notation  $\zeta(2)=\pi^2/6$ and introduced the
 label ``SQCD'' to specify that the strong coupling has to be evaluated
 within the full theory, i.e.\ the SQCD with $n_f=n_s=6$ active flavours.

The presence of the terms proportional with the parameter $X_b$ is a
manifestation of the supersymmetry breaking. They are generated
by the Yukawa interaction between left- and right-handed bottom squarks
and the CP-neutral Higgs fields. 
 Their contribution to the decoupling
coefficient of the bottom-quark mass can be related through the Low
Energy Theorem~\cite{Ellis:1975ap} to the decay rate of the Higgs boson to $b\bar{b}$
pairs. To one-loop order, the  $X_b$-term of
Eq.~(\ref{eq::zetam1l}) coincides with the SQCD corrections to the decay
rate $\phi\to b\bar{b}$\cite{Guasch:2003cv}. To higher orders, the relation
between the two parameters becomes more involved. 
\\
The Yukawa-coupling induced contributions attracted a lot of attention in
the past years, due to the fact that they  are the dominant
corrections for large values of $\tan \beta$. They may in general become
comparable with the tree-level bottom-quark mass. The resummation of the
one-loop corrections was performed in Ref.~\cite{Carena:1999py}.

\subsection{\label{sec::twoloop} Two-loop calculation}
The analytical two-loop results for the decoupling coefficients are
too lengthy to be displayed here. They are available in {\tt
  MATHEMATICA} format from
http://www-ttp.particle.uni-karlsruhe.de/Progdata/ttp08-25. 
Instead, we present the two-loop results for three special cases of the
hierarchy among 
the heavy particle masses.
Before displaying the analytical results, let us notice the absence of
contributions of the form $\alpha_s^2 X_b^2$ to $\zeta_{m_b}$, in
accordance with 
Refs.~\cite{Carena:1999py, Guasch:2003cv}. They are suppressed by a factor
$m_b/\tilde{M}$, that we neglect in the ET formalism.
% For the shake of
%compactness of the analytical formulae, we also do not
%present the two-loop contributions to $\zeta_{e,q}$. They are available
%upon request.
%- }}}

\subsubsection{Scenario A}
We consider first the case of all supersymmetric particles having masses
of the same order of magnitude and being much heavier than the top-quark
\[
\begin{split}
&m_{\tilde{u}} = \ldots = m_{\tilde{b}} = m_{\tilde{t}} =
  m_{\tilde{g}}=\tilde{M}\quad \gg\quad 
m_t\nonumber\\ 
& \alpha_s^{(5)} = \zeta_{s}^{\tilde{M}}\, \alpha_s^{\rm{ (SQCD)}} \,,\qquad 
m_b^{(5)} = \zeta_{m_b}^{\tilde{M}}\, m_b^{\rm{ (SQCD)}}\,.
\end{split}
\]
$\zeta_{s}^{\tilde{M}},\,\zeta_{m_b}^{\tilde{M}} $ are functions of the
supersymmetric mass $\tilde{M}$ and the top-quark mass $m_t$, the
soft SUSY breaking
parameters $X_q$, $q=b,t$, the strong
coupling constant  in the full theory $ \alpha_s^{\rm{ (SQCD)}}$  and the
decoupling scale $\mu$. The superscript $(5)$ indicates that the
parameters are defined in QCD with $n_l=5$ light quarks. In addition to
the notations introduced in 
Eq.~(\ref{eq::abbv}), the following  abbreviation will be used
\begin{eqnarray}
L_{\tilde{M}}=\ln\frac{\mu^2}{\tilde{M}^2}\,.
\end{eqnarray}
 The two-loop result for the decoupling coefficient of $\alpha_s$ in
 case of a degenerate SUSY mass spectrum is
 known~\cite{Harlander:2005wm}, however we 
 give it here for completeness  
{\allowdisplaybreaks
\begin{align}
 \zeta_{s}^{\tilde{M}}&= 1+\frac{\alpha _s^{\rm\tiny{ (SQCD)}}}{\pi } 
\Bigg[ C_A \left(-\frac{1}{6} L_{\tilde M}\right)
+\Bigg(-L_{\tilde M}-\frac{L_t}{3}\Bigg) T_F\Bigg]
+ \Bigg(\frac{\alpha _s^{\rm\tiny{ (SQCD)}}}{\pi }\Bigg)^2 \Bigg\{C_A^2
\Bigg(-\frac{85}{288}\nonumber\\
&-\frac{L_{\tilde M}}{3}+\frac{L_{\tilde
    M}^2}{36}\Bigg)  
%\nonumber  \\ &
+ C_F T_F \Bigg[-\frac{31}{16}-\frac{3 L_{\tilde
    M}}{2}-\frac{L_t}{4}-\frac{m_t^2}{8 \tilde{M}^2}+\frac{\pi
  m_t^3}{12 \tilde{M}^3}+\Bigg(-\frac{17}{150}-\frac{3L_{\tilde M}}{40 
}\nonumber\\
&+\frac{3L_t}{40 }\Bigg) \frac{m_t^4}{\tilde{M}^4} \Bigg]
+
C_A T_F \Bigg[\frac{41}{36}+L_{\tilde M}+\frac{L_{\tilde
    M}^2}{3}-\frac{5L_t}{12}+\frac{L_{\tilde M} L_t}{9} 
+\Bigg(\frac{1}{8 }+\frac{L_{\tilde M}}{4 }-\frac{L_t}{4 }\Bigg)
\frac{m_t^2}{m_S^2} \nonumber \\
&-\frac{\pi  m_t^3}{6 \tilde{M}^3}
 +\Bigg(\frac{19}{144}+\frac{ L_{\tilde M}}{24}-\frac{ L_t}{24}\Bigg)
\frac{m_t^4}{\tilde{M}^4} \Bigg]
+T_F^2\Bigg(L_{\tilde M}^2+\frac{2
  L_{\tilde M} L_t}{3}+\frac{L_t^2}{9}\Bigg) \nonumber \\
& + {\cal O}\left(
\frac{m_t^5}{\tilde{M}^5} \right) \Bigg\}\,,
\label{eq::degas}
\\
\zeta_{m_b}^{\tilde{M}} &= 1-\frac{\alpha _s^{\rm\tiny{ (SQCD)}}}{\pi }
C_F\left[\frac{ L_{\tilde M} }{4} + 
\frac{X_b}{4 {\tilde M} }\right]+\left(\frac{\alpha _s^{\rm\tiny{
      (SQCD)}}}{\pi }\right)^2 C_F\Bigg\{   
-C_A\left(\frac{65}{1152} + \frac{43 L_{\tilde M} }{96}  +
  \frac{L_{\tilde M}^2}{32} \right)
\nonumber\\
&+ C_F\left(-\frac{99}{128} - \frac{7 L_{\tilde M}}{32} +
\frac{L_{\tilde M}^2}{32} \right) 
+T_F\bigg[\frac{197}{72} - L_{\tilde M} + \frac{3 L_{\tilde M}^2}{4} -
  \frac{\lnMt}{12} + \frac{\lnMt^2}{8} 
\nonumber\\
&+\left( \frac{7}{144} + \frac{L_{\tilde M}}{12} -
\frac{\lnMt}{12}\right)\frac{m_t^2}{\tilde{M}^2} 
-\frac{\pi}{12}\frac{m_t^3}{\tilde{M}^3}
+\left(\frac{53}{600} - \frac{7 L_{\tilde M}}{160} + \frac{7
    \lnMt}{160}\right)\frac{m_t^4}{\tilde{M}^4} 
\bigg]
\nonumber\\
&+
\frac{X_b}{\tilde{M}}\Bigg[
-C_A\left( \frac{1}{16} + \frac{3 L_{\tilde M}}{16} \right)-
C_F\left(\frac{1}{4} - \frac{3 L_{\tilde M}}{16} \right)
+T_F\bigg[\frac{-3}{4} + \frac{3 L_{\tilde M} }{4}  
\nonumber\\
&+
\left(\frac{-7}{72} - \frac{L_{\tilde M}}{24} +
\frac{\lnMt}{24}\right)\frac{m_t^2}{\tilde{M}^2} 
+\frac{\pi}{24}\frac{m_t^3}{\tilde{M}^3}
+\left(-\frac{17}{450} + \frac{L_{\tilde M}}{240} -
\frac{\lnMt}{240}\right)\frac{m_t^4}{\tilde{M}^4} 
\bigg]
\Bigg]
\nonumber\\
&+
\frac{X_t}{\tilde{M}}T_F\Bigg[
\left( \frac{-5}{72} + \frac{L_{\tilde M}}{12} - \frac{\lnMt}{12}
\right)\frac{m_t^2}{\tilde{M}^2} 
-\frac{\pi}{24}\frac{m_t^3}{\tilde{M}^3}
+\left(\frac{23}{450} - \frac{L_{\tilde M}}{60} + \frac{\lnMt}{60}
\right)\frac{m_t^4}{\tilde{M}^4} 
\Bigg]
\nonumber\\
&+
\frac{X_t X_b}{\tilde{M}^2} T_F\Bigg[
\left( \frac{1}{72} - \frac{L_{\tilde M}}{24} + \frac{\lnMt}{24}
\right)\frac{m_t^2}{\tilde{M}^2} 
+\frac{\pi}{48}\frac{m_t^3}{\tilde{M}^3}
+\left(\frac{-59}{1800} - \frac{L_{\tilde M}}{120} + \frac{\lnMt}{120}
\right)\frac{m_t^4}{\tilde{M}^4} \Bigg]
\nonumber\\
&+ 
{\cal O}\left(
\frac{m_t^5}{\tilde{M}^5} \right) 
\Bigg\}\,.
\end{align}
}

Let us point out that, according to Eq.~(\ref{eq::mixb}) the assumption of
degenerate top-squark masses 
can be  materialized only if $X_t\to 0$, due to the heavy top-quark
mass. We display, however, for completeness  the full result. Further on, the
hypothesis of equal top- and bottom-squark masses is inconsistent with
the $SU(2)$ invariance  of the $\tilde{t}/\tilde{b}$ isodublet imposed
in models like mSUGRA. 

\subsubsection{Scenario B}
In the following, we discuss the possibility that the gluino is the heaviest
supersymmetric particle and the squarks have equal masses, much heavier
than  that of the top-quark
\[
\begin{split}
&m_{\tilde{g}} \quad \gg\quad  m_{\tilde{u}} = \ldots = m_{\tilde{b}} =
m_{\tilde{t}} =  m_{\tilde{q}}\quad \gg\quad 
m_t\nonumber\\ 
& \alpha_s^{(5)} = \zeta_{s}^{\tilde{g}}\, \alpha_s^{\rm{ (SQCD)}} \,,\qquad 
m_b^{(5)} = \zeta_{m_b}^{\tilde{g}}\, m_b^{\rm{ (SQCD)}}\,.
\end{split}
\]
The two-loop results read
{\allowdisplaybreaks
\begin{align}
 \zeta_{s}^{\tilde{g}}&=
1+\frac{\alpha _s^{\rm\tiny{ (SQCD)}}}{\pi } \Bigg[C_A\left(-\frac{1}{6} 
\lgl\right)+\Bigg(-\frac{L_t}{3}-\lsq\Bigg) T_F\Bigg]+ 
\Bigg(\frac{\alpha _s^{\rm\tiny{ (SQCD)}}}{\pi }\Bigg)^2
\Bigg\{C_A^2
  \Bigg(-\frac{85}{288}-\frac{\lgl}{3}\nonumber\\
 &
+\frac{\lgl^2}{36}\Bigg) 
+
T_F^2\Bigg(\frac{L_t^2}{9}+\frac{2}{3} L_t \lsq+\lsq^2\Bigg)
+C_F T_F \Bigg[-\frac{19}{16}-\frac{L_t}{4}+\frac{3 \lgl}{2}-3 \lsq 
+\Bigg(\frac{1}{6}+\lgl\nonumber\\ 
&
-\lsq\Bigg) \frac{\msq^2}{\mg^2}
+
\Bigg(\frac{1}{24}+\lgl-\lsq\Bigg)\frac{ \msq^4}{\mg^4}
+\Bigg(-\frac{5}{9}+\frac{L_t}{3}-\frac{\lgl}{3}\Bigg) \frac{
m_t^2}{\mg^2}
+\Bigg(-\frac{9}{8}+\frac{5 L_t}{4}-\frac{5 \lgl}{4}\Bigg)
\frac{m_t^4}{\mg^4} 
\nonumber\\
 &+\frac{m_t^2 \msq^2}{\mg^4} \Bigg(-\frac{41}{36}+\frac{11
  L_t}{6}-\frac{23 
\lgl}{6}+L_t \lgl-\lgl^2+2 \lsq-L_t \lsq+\lgl \lsq-2 \zeta
_2\Bigg)\Bigg]
\nonumber\\
 &+C_A T_F \Bigg[\frac{95}{36}-\frac{5 L_t}{12}+\frac{3
  \lgl}{2}+\frac{1}{9} L_t \lgl-\frac{\lsq}{2}+\frac{1}{3} \lgl \lsq
+\Bigg(-\frac{3 \lgl}{2}+\frac{3
  \lsq}{2}\Bigg)\frac{\msq^2}{\mg^2}
\nonumber\\
 &+
\Bigg(-\frac{3}{2}+\frac{3 \lgl}{2}-\frac{3 
\lsq}{2}\Bigg) \frac{\msq^4}{\mg^4}
+\Bigg(\frac{1}{2}-\frac{L_t}{4}+\frac{\lgl}{4}\Bigg)\frac{
  m_t^2}{\mg^2}
+\Bigg(\frac{1}{4}-\frac{L_t}{4}+\frac{\lgl}{4}\Bigg)
\frac{m_t^4}{\mg^4}
\nonumber\\
 &+\frac{m_t^2 \msq^2}{\mg^4} \Bigg(\frac{1}{2}-\frac{3
  L_t}{4}+\frac{3 \lgl}{2}-\frac{1}{2} L_t
\lgl+\frac{\lgl^2}{2}-\frac{3\lsq}{4}+\frac{1}{2} L_t \lsq-\frac{1}{2}
\lgl \lsq 
+\zeta _2\Bigg)\Bigg] \nonumber\\
& + {\cal O}\left( \frac{\msq^6}{\mg^6},\frac{\msq^4
    \mt^2}{\mg^6},\frac{\msq^2 \mt^4}{\mg^6}, \frac{\mt^6}{\mg^6}
\right)\bigg\}\,, 
\end{align}
}
{\allowdisplaybreaks
\begin{align}
\zeta_{m_b}^{\tilde{g}} &= 1+ \frac{\alpha _s^{\rm\tiny{ (SQCD)}}}{\pi }
C_F\bigg\{-\frac{3}{8} - \frac{\lnMgl}{4}
+\left(-\frac{1}{4} - \frac{\lnMgl}{2} + \frac{\lsq}{2}
\right)\frac{\msq^2}{\mg^2}
+\left(-\frac{1}{4} - \frac{3 \lnMgl}{4} + \frac{3
    \lsq}{4}\right)\frac{\msq^4}{\mg^4}
 \nonumber\\
&+\frac{X_b}{\mg}\left[
\frac{1}{2} + \frac{\lnMgl}{2} - \frac{\lsq}{2}+\left(\frac{1}{2} + \lnMgl -
\lsq\right)\frac{\msq^2}{\mg^2}
+ \left(\frac{1}{2} + \frac{3\lnMgl}{2} - \frac{3\lsq}{2}
\right)\frac{\msq^4}{\mg^4} 
\right]
 \bigg\}
 \nonumber\\
&+\left(\frac{\alpha _s^{\rm\tiny{ (SQCD)}}}{\pi }\right)^2 C_F\bigg\{
C_A\bigg[ -\frac{209}{1152} - \frac{35 \lnMgl}{48} - \frac{\lnMgl^2}{32}
- \frac{\zeta(2)}{8}
+\bigg(-\frac{5}{16} - \frac{3\lnMgl}{2} - \frac{5\lnMgl^2}{8} 
\nonumber\\
&
+\frac{21\lsq}{16} 
+ \frac{7\lnMgl\lsq}{8} -  \frac{\lsq^2}{4}
  - \frac{5\zeta(2)}{8} \bigg)\frac{\msq^2}{\mg^2}
+\bigg(
\frac{7}{32} - \frac{11\lnMgl}{4} - \frac{37\lnMgl^2}{32} +
\frac{41\lsq}{16} + \frac{7\lnMgl\lsq}{4} -  
 \frac{19\lsq^2}{32} 
\nonumber\\
&- \frac{5\zeta(2)}{8}
\bigg)\frac{\msq^4}{\mg^4}
\bigg]+C_F\bigg[
-\frac{221}{128} - \frac{\lnMgl}{8} + \frac{\lnMgl^2}{32}
-\frac{\zeta(2)}{16}
+\bigg(\frac{9}{16} + \frac{5\lnMgl^2}{8} + \frac{\lsq}{16} -
\frac{9\lnMgl\lsq}{8} 
\nonumber\\
&+
  \frac{\lsq^2}{2} + \frac{\zeta(2)}{2}\bigg)\frac{\msq^2}{\mg^2} 
+\bigg(\frac{31}{192} + \frac{35\lnMgl}{32} + \frac{17\lnMgl^2}{16} -
\frac{33\lsq}{32} -  
 \frac{31\lnMgl\lsq}{16} + \frac{7\lsq^2}{8}
+\frac{\zeta(2)}{8}
\bigg)\frac{\msq^4}{\mg^4} \bigg]
\nonumber\\
& + T_F\bigg[
\frac{139}{36} + \frac{3\lnMgl}{2} + \frac{3\lnMgl^2}{8} -
\frac{11\lsq}{8}  +  
\frac{3\lsq^2}{8} - \frac{\lnMt}{12} + \frac{\lnMt^2}{8} +
\frac{3\zeta(2)}{4}
\nonumber\\
&+\bigg(
3 + \frac{9\lnMgl}{2}  +3 \lnMgl^2- \frac{15\lsq}{4} - \frac{9\lnMgl\lsq}{2} + 
 \frac{3\lsq^2}{2} \bigg)\frac{\msq^2}{\mg^2}
+\bigg(-\frac{1}{2} - \frac{\lnMgl}{4} + \frac{\lsq}{4}
\bigg)\frac{m_t^2}{\mg^2}
\nonumber\\
&-\bigg(
1 + \frac{\lnMgl}{2} - \frac{3\lnMgl^2}{8} - \frac{\lsq}{2} + \frac{3\lnMgl\lsq}{4}
- \frac{3\lsq^2}{8} - \frac{3\zeta(2)}{4}
\bigg)\frac{\msq^2 m_t^2}{\mg^4}
\nonumber\\
&+
\bigg(-\frac{5}{72}+\frac{\lsq}{24}-\frac{\lnMt}{24}\bigg)\frac{
  m_t^4}{\msq^2 \mg^2}+\bigg(
-\frac{155}{96} - \frac{3\lnMgl}{16} + \frac{3\lnMgl^2}{16} +
\frac{7\lsq}{8} - \frac{3\lsq^2}{16} -  
 \frac{11\lnMt}{16} 
\nonumber\\
&- \frac{3\lnMgl\lnMt}{8} + \frac{3\lsq\lnMt}{8} + \frac{3\zeta(2)}{8}
\bigg)\frac{m_t^4}{\mg^4}
+\bigg(\frac{69}{16} + \frac{63\lnMgl}{8} + 3\lnMgl^2 - \frac{57\lsq}{8}
- \frac{15\lnMgl\lsq}{4} +  
 \frac{3\lsq^2}{4}
\bigg)\frac{\msq^4}{\mg^4}\bigg]
\nonumber\\
&+\frac{X_b}{\mg}\bigg[
C_A\bigg[\frac{7}{8} + \lnMgl + \frac{3\lnMgl^2}{8} - \frac{5\lsq}{8} -
\frac{3\lnMgl\lsq}{8} - \frac{\zeta(2)}{4}
+\bigg(\frac{9}{8} + \frac{17\lnMgl}{8} + \frac{11\lnMgl^2}{8} -
  \frac{7\lsq}{4} 
\nonumber\\
&- 2\lnMgl\lsq + 
 \frac{5\lsq^2}{8} - \frac{\zeta(2)}{4} \bigg)\frac{\msq^2}{\mg^2} 
+\bigg(
\frac{19}{16} + \frac{27\lnMgl}{8} + \frac{23\lnMgl^2}{8} - 3\lsq -
\frac{37\lnMgl\lsq}{8} +  
 \frac{7\lsq^2}{4} 
\nonumber\\
&- \frac{\zeta(2)}{4} \bigg)\frac{\msq^4}{\mg^4} 
\bigg]
+C_F\bigg[\frac{1}{16} - \frac{3\lnMgl}{16} - \frac{3\lnMgl^2}{8} -
\frac{3\lsq}{16} + \frac{3\lnMgl\lsq}{8} +  \frac{\zeta(2)}{2} 
+\bigg(-\frac{9}{16} - \frac{5\lnMgl}{24} 
\nonumber\\
&- \frac{3\lnMgl^2}{2} -
\frac{\lsq}{6} + \frac{9\lnMgl\lsq}{4} -   \frac{3\lsq^2}{4} +
\frac{\zeta(2)}{2} 
\bigg)\frac{\msq^2}{\mg^2} 
+\bigg(-\frac{13}{16} - \frac{3\lnMgl}{4} - \frac{13\lnMgl^2}{4} +
\frac{3\lsq}{8} 
\nonumber\\
&+ \frac{43\lnMgl\lsq}{8} - \frac{17\lsq^2}{8} + \frac{\zeta(2)}{2}
\bigg)\frac{\msq^4}{\mg^4} 
\bigg]
+T_F\bigg[
-\frac{3}{2}(2 + \lnMgl)(1 + \lnMgl - \lsq)
+\bigg(-9 - \frac{15\lnMgl}{2} 
\nonumber\\
&- 6\lnMgl^2 + 6\lsq + 9\lnMgl\lsq - 3\lsq^2
\bigg)\frac{\msq^2}{\mg^2}
+\bigg(\frac{3}{2}+\frac{\lnMgl}{2}-\frac{\lsq}{2}
\bigg)\frac{m_t^2}{\mg^2}
+\bigg(-\frac{33}{4} - \frac{63\lnMgl}{4} - 6\lnMgl^2 
\nonumber\\
&+ \frac{57\lsq}{4} + \frac{15\lnMgl\lsq}{2} - 
 \frac{3\lsq^2}{2}\bigg)\frac{\msq^4}{\mg^4}
+\bigg(\frac{5}{18} - \frac{\lsq}{6} +
\frac{\lnMt}{6}\bigg)\frac{m_t^4}{\msq^2\mg^2}
\nonumber\\
&+\bigg(\frac{5}{2} + \frac{3\lnMgl}{2} - \frac{\lnMgl^2}{2} -
\frac{3\lsq}{2} + \lnMgl\lsq - \frac{\lsq^2}{2} -
2\zeta(2)\bigg)\frac{m_t^2 \msq^2}{\mg^4}
\nonumber\\
&+\bigg(\frac{301}{72} + \frac{5\lnMgl}{8} 
- \frac{\lnMgl^2}{4} 
-\frac{53\lsq}{24} - \frac{\lnMgl\lsq}{4} + \frac{\lsq^2}{2} +
\frac{19\lnMt}{12} + \frac{3\lnMgl\lnMt}{4} 
- \frac{3\lsq\lnMt}{4} - \zeta(2) 
\bigg)\frac{\mt^4}{\mg^4}
\bigg]
\bigg]
\nonumber\\
&+\frac{X_t}{\mg}T_F\bigg[\bigg(\frac{1}{2} - \lnMgl - \frac{\lnMgl^2}{4} +
\lsq + \frac{\lnMgl\lsq}{2} - \frac{\lsq^2}{4} - \frac{\zeta(2)}{2}
\bigg)\frac{m_t^2}{\mg^2} 
+\bigg(
1 - \frac{7\lnMgl}{2} - 2\lnMgl^2 + \frac{7\lsq}{2} 
\nonumber\\
&+ 4\lnMgl\lsq - 2\lsq^2 -  2\zeta(2)
\bigg)\frac{m_t^2 \msq^2}{\mg^4}+\bigg(\frac{4}{9} - \frac{\lsq}{6} +
\frac{\lnMt}{6}\bigg)\frac{m_t^4}{\mg^2 \msq^2} +\bigg(\frac{17}{9} -
3\lnMgl - \lnMgl^2 + \frac{\lsq}{6} 
\nonumber\\
&+ \lnMgl\lsq + \frac{17\lnMt}{6} + 
 \lnMgl\lnMt - \lsq\lnMt - 2\zeta(2)\bigg)\frac{m_t^4}{\mg^4 }
\bigg]
+\frac{X_t X_b}{\mg^2}T_F\bigg[-\frac{1}{2}\frac{m_t^2}{\msq^2 }
+
\bigg(-\frac{3}{2} + \frac{5\lnMgl}{2} 
\nonumber\\
&
+ \frac{5\lnMgl^2}{4} -
\frac{5\lsq}{2} - \frac{5\lnMgl\lsq}{2} +  
 \frac{5\lsq^2}{4} + \frac{3\zeta(2)}{2}\bigg)\frac{m_t^2}{\mg^2 }
\bigg]
%\nonumber\\
%&
+{\cal O}\left( \frac{\msq^6}{\mg^6},\frac{\msq^4
    \mt^2}{\mg^6},\frac{\msq^2 \mt^4}{\mg^6},
  \frac{\mt^6}{\mg^6},\frac{X_t X_b\mt^4}{\mg^6}
\right)\bigg\}\,. 
\end{align}
}

\subsubsection{Scenario C}
Finally, we make the assumption that all squark masses are degenerate
and are much heavier than the gluino and top masses  
\[
\begin{split}
& m_{\tilde{u}} = \ldots = m_{\tilde{b}} =
m_{\tilde{t}} =  m_{\tilde{q}}\quad \gg\quad m_{\tilde{g}} \quad \gg\quad
m_t\nonumber\\ 
& \alpha_s^{(5)} = \zeta_{s}^{\tilde{q}}\, \alpha_s^{\rm{ (SQCD)}} \,,\qquad 
m_b^{(5)} = \zeta_{m_b}^{\tilde{q}}\, m_b^{\rm{ (SQCD)}}\,.
\end{split}
\]
The decoupling coefficients are given by 
{\allowdisplaybreaks
\begin{eqnarray}
 \zeta _{s}^{\tilde{q}}&=& 1+\frac{\alpha _s^{\rm{ (SQCD)}}}{\pi }
\Bigg[-\frac{1}{6} C_A \lgl+\Bigg(-\frac{L_t}{3}-\lsq\Bigg)
T_F\Bigg]+\Bigg(\frac{\alpha _s^{\rm{ (SQCD)}}}{\pi }\Bigg)^2 \Bigg\{C_A^2
  \Bigg(-\frac{85}{288}-\frac{\lgl}{3}
\nonumber \\&+&
\frac{\lgl^2}{36}\Bigg)
+C_F T_F \Bigg[-\frac{7}{16}-\frac{L_t}{4}-\frac{3
\lsq}{2}+\Bigg(-\frac{1}{9}-\frac{L_t}{12}+\frac{\lsq}{12}\Bigg)
\frac{m_t^4}{\msq^4}+\Bigg(-\frac{1}{9}-\frac{L_t}{3}
\nonumber\\
 &+&
\frac{\lsq}{3}\Bigg) 
\frac{m_t^2 \mg^2}{\msq^4} 
+\Bigg(-\frac{3}{4}-\frac{3 \lgl}{2}+\frac{3 \lsq}{2}\Bigg)
\frac{\mg^4}{\msq^4}+\Bigg(\frac{1}{18}+\frac{L_t}{6}-\frac{\lsq}{6}\Bigg) 
\frac{m_t^2}{\msq^2}\Bigg]
+T_F^2 \Bigg(\frac{L_t^2}{9}\nonumber\\ &+&
\frac{2}{3} L_t \lsq+\lsq^2\Bigg)
+C_A T_F \Bigg[\frac{41}{36}-\frac{5 L_t}{12}+\frac{1}{9}
L_t \lgl+\lsq+\frac{1}{3} \lgl
\lsq+\Bigg(-\frac{L_t}{4}+\frac{\lsq}{4}\Bigg)
\frac{m_t^2}{\msq^2}
\nonumber\\ 
&+&
\Bigg(\frac{1}{3}-\frac{\lgl}{2}+\frac{\lsq}{2}\Bigg) 
\frac{\mg^2}{\msq^2}
+\Bigg(-\frac{L_t}{4}+\frac{\lsq}{4}\Bigg)
\frac{m_t^4}{\msq^4}+\Bigg(\frac{1}{12}-\frac{\lgl}{2}+\frac{\lsq}{2}\Bigg) 
\frac{\mg^4}{\msq^4} 
\nonumber\\
 &+&\frac{m_t^2 \mg^2}{\msq^4} \Bigg(\frac{7}{18}-\frac{3
  L_t}{4}-\frac{13 \lgl}{12}+\frac{1}{2} L_t \lgl+\frac{11
  \lsq}{6}-\frac{1}{2} 
L_t \lsq-\frac{1}{2} \lgl \lsq+\frac{\lsq^2}{2}+\zeta
_2\Bigg)\Bigg] 
\nonumber\\
& +& {\cal O}\left( \frac{\mg^6}{\msq^6},\frac{\mg^4
    \mt^2}{\msq^6},\frac{\mg^2 \mt^4}{\msq^6}, \frac{\mt^6}{\msq^6}
\right)\Bigg\}\,, 
\end{eqnarray}
}
{\allowdisplaybreaks
\begin{eqnarray}
\zeta_{m_b}^{\tilde{q}} &=& 1 + \frac{\alpha _s^{\rm{ (SQCD)}}}{\pi } C_F
\Bigg\{-\frac{1}{8}-\frac{\lsq}{4}+\frac{\mg^2}{4\msq^2} 
+\Bigg(\frac{1}{4}-\frac{\lgl}{4}+\frac{\lsq}{4}\Bigg)\frac{\mg^4}{\msq^4}
+ \frac{X_b}{\msq}\Bigg[-\frac{\mg}{2\msq}\nonumber\\&+&
\Bigg(-\frac{1}{2}+\frac{\lgl}{2}-\frac{\lsq}{2}\Bigg)
\frac{\mg^3}{\msq^3}\Bigg]\Bigg\}  
 +\left(\frac{\alpha _s^{\rm{ (SQCD)}}}{\pi }\right)^2 C_F \Bigg\{
 C_A
 \Bigg[\frac{295}{1152}-\frac{\lgl}{24}+\frac{\lgl^2}{16}-\frac{\lsq}{2}
-\frac{3\lsq^2}{32}\nonumber\\&-&\frac{\zeta(2)}{2}  
+\Bigg(\frac{15}{16}+\frac{\lgl}{2}-\frac{5\lsq}{16}
-\frac{5\zeta(2)}{8}\Bigg)\frac{\mg^2}{\msq^2} 
 +\Bigg(\frac{19}{16}+\frac{3\lgl}{16}+\frac{7\lgl^2}{32}
+\frac{13\lsq^2}{32}-\frac{5\lgl\lsq}{8}
\nonumber\\&-&\frac{5\zeta(2)}{8}\Bigg)\frac{\mg^4}{\msq^4}\Bigg]
 + C_F \Bigg[-\frac{205}{128}-\frac{3\lsq}{16}+\frac{\lsq^2}{32}
+\frac{15\zeta(2)}{16}
 +\Bigg(-\frac{3}{2}-\frac{\lsq}{16}
+\frac{5\zeta(2)}{4}\Bigg)\frac{\mg^2}{\msq^2}
 \nonumber\\&+&\Bigg(-\frac{187}{64}-\frac{39\lgl}{32}
-\frac{7\lgl^2}{8}+\frac{37\lsq}{32} 
 +\frac{29\lsq\lgl}{16}-\frac{15\lsq^2}{16}
+\frac{13\zeta(2)}{8}\Bigg)\frac{\mg^4}{\msq^4}\Bigg]
 \nonumber\\&+& T_F \Bigg[\frac{28}{9}-\frac{\lt}{12}+\frac{\lt^2}{8}
-\frac{5\lsq}{8}+\frac{3\lsq^2}{4}-\frac{3\zeta(2)}{4}
 +\Bigg(\frac{3}{4}-\frac{3\lsq}{4}\Bigg)\frac{\mg^2}{\msq^2}
 +\Bigg(-\frac{1}{2}+\frac{\zeta(2)}{4}\Bigg)\frac{\mt^2}{\msq^2}
 \nonumber\\&+&\Bigg(-\frac{13}{16}+\frac{3\lgl}{8}-\frac{9\lsq}{8}
+\frac{3\lgl\lsq}{4}-\frac{3\lsq^2}{4}\Bigg)\frac{\mg^4}{\msq^4} 
 +\Bigg(-\frac{5}{2}+\frac{3\zeta(2)}{2}\Bigg)\frac{\mt^2\mg^2}{\msq^4}
 \nonumber\\&+&\Bigg(\frac{61}{288}-\frac{\zeta(2)}{8}+\frac{\lt}{48}
-\frac{\lsq}{48}\Bigg)\frac{\mt^4}{\msq^4}
 \Bigg]
 + \frac{X_b}{\msq} \Bigg[ C_A \Bigg[
 \Bigg(-\frac{7}{8}+\frac{3\lgl}{8}-\frac{3\lsq}{4}
-\frac{\zeta(2)}{4}\Bigg)\frac{\mg}{\msq}
 \nonumber\\&+&\Bigg(-\frac{5}{8}+2\lgl-\frac{19\lsq}{8}+\frac{3\lgl\lsq}{8}
-\frac{3\lsq^2}{8}-\frac{\zeta(2)}{4}\Bigg)\frac{\mg^3}{\msq^3}
 \Bigg]
 + C_F \Bigg[
 \Bigg(-\frac{3}{16}+\frac{3\lsq}{8}
\nonumber\\&+&\frac{\zeta(2)}{2}\Bigg)\frac{\mg}{\msq}
 +\Bigg(-\frac{13}{16}-\frac{13\lgl}{16}+\frac{19\lsq}{16}-\frac{3\lgl\lsq}{8}
+\frac{3\lsq^2}{8}+\frac{\zeta(2)}{2}\Bigg)\frac{\mg^3}{\msq^3}
 \Bigg]
 \nonumber\\&+& T_F \Bigg[
 \Bigg(-\frac{3}{4}+\frac{3\lsq}{2}\Bigg)\frac{\mg}{\msq}
 +\Bigg(\frac{7}{4}-\frac{3\lgl}{4}+\frac{9\lsq}{4}-\frac{3\lgl\lsq}{2}
+\frac{3\lsq^2}{2}\Bigg)\frac{\mg^3}{\msq^3}
 \nonumber\\&+&\Bigg(\frac{13}{4}-2\zeta(2)\Bigg)\frac{\mt^2\mg}{\msq^3}
 \Bigg] \Bigg]
 - \frac{X_t}{\msq} T_F \frac{\zeta(2)}{2} \frac{\mt^2\mg}{\msq^3}
 + \frac{X_tX_b}{\msq^2} T_F \Bigg[
 \frac{\zeta(2)}{2} \frac{\mt^2}{\msq^2}\nonumber\\&+&
 \Bigg(\frac{3}{4}-3\lgl+
3\lsq+3\zeta(2)\Bigg)\frac{\mt^2\mg^2}{\msq^4}
 +\Bigg(\frac{5}{9}-\frac{2\lt}{3}+\frac{2\lsq}{3}\Bigg)\frac{\mt^4}{\msq^4}
 \Bigg]
\nonumber\\
&+&{\cal O}\left( \frac{\mg^6}{\msq^6},\frac{\mg^4
    \mt^2}{\msq^6},\frac{\mg^2 \mt^4}{\msq^6},
  \frac{\mt^6}{\msq^6},\frac{X_b \mg^5}{\msq^6}, \frac{X_t \mg^5}{\msq^6}
\right)  \Bigg\}
\,. 
\end{eqnarray}
}

We displayed in the previous expressions only the first
three terms of the Taylor expansions in the mass ratios. To get an
idea about the convergence of the perturbative series we fix the following input
parameters: $\mt=172.4$~GeV, $\alpha_s^{\rm{ (SQCD)}}=0.120$,
$\msq=500$~GeV, $X_q=-4000$~GeV, $X_t=-400$~GeV and let $\mg$
vary. Even for
$\mg/\msq= 0.5,$ and
$2$ the approximations given above agree with the exact results with an
accuracy better than  1\%. For the case of degenerate SUSY masses, {\it
  i.e\ } $\mg/\msq=1$ the accuracy is  even  below the per-mille level.

%- }}}

%-{{{numerics
\section{\label{sec::num} Numerical results}
In this Section we discuss the numerical impact of the two-loop
calculations we  presented above.  A first
phenomenological application is the prediction of
the strong coupling and the running bottom-quark  mass at high-energy scales
like $\tilde{M}= 1$~TeV or $\mu_{\rm GUT}=10^{16}$~GeV, starting from
their low-energy values determined experimentally. 

For the energy evolution of the two parameters, we follow the method
proposed in Ref.~\cite{Harlander:2007wh}: first, we compute
$\alpha_s^{(5)}(\mu_{\rm dec})$ and $m_b^{(5)}(\mu_{\rm dec})$ from 
$\alpha_s^{(5)}(M_Z)$ and $m_b^{(5)}(m_b)$, respectively, using the
corresponding $i$-loop SM RGEs~\cite{Steinhauser:2002rq}. Here $\mu_{\rm
  dec}$ denotes the energy 
scale at which the heavy particles are supposed to become active, {\it
  i.e.\ }
 the scale where the matching between the SM and the MSSM is
performed. As pointed out in  previous
works~\cite{Ferreira:1996ug}, one can avoid part of the 
complications related with the occurrence of the {\it 
  evanescent couplings}, performing the change of the regularization
scheme  from the \msbar{}
to the \drbar{}  scheme at the same
scale. Nevertheless, one cannot avoid the occurrence of the evanescent
coupling $\alpha_e$  in the 
\msbar{}-\drbar{}  relation for the bottom-quark mass. It has
to be determined iteratively from the knowledge of the strong coupling
at the matching scale.  
 For consistency, the  $i$-loop running parameters have to be folded with
 $(i-1)$-loop conversion and decoupling relations. 
 Above the decoupling
scale, the energy dependence of the running parameters is governed by
the $i$-loop  MSSM  RGEs~\cite{Ferreira:1996ug}. We solved numerically
the system of coupled 
differential equations arising from the two sets of RGEs, and implemented
this procedure for $i =1,2,3$.

\begin{figure}[t]
\epsfig{figure=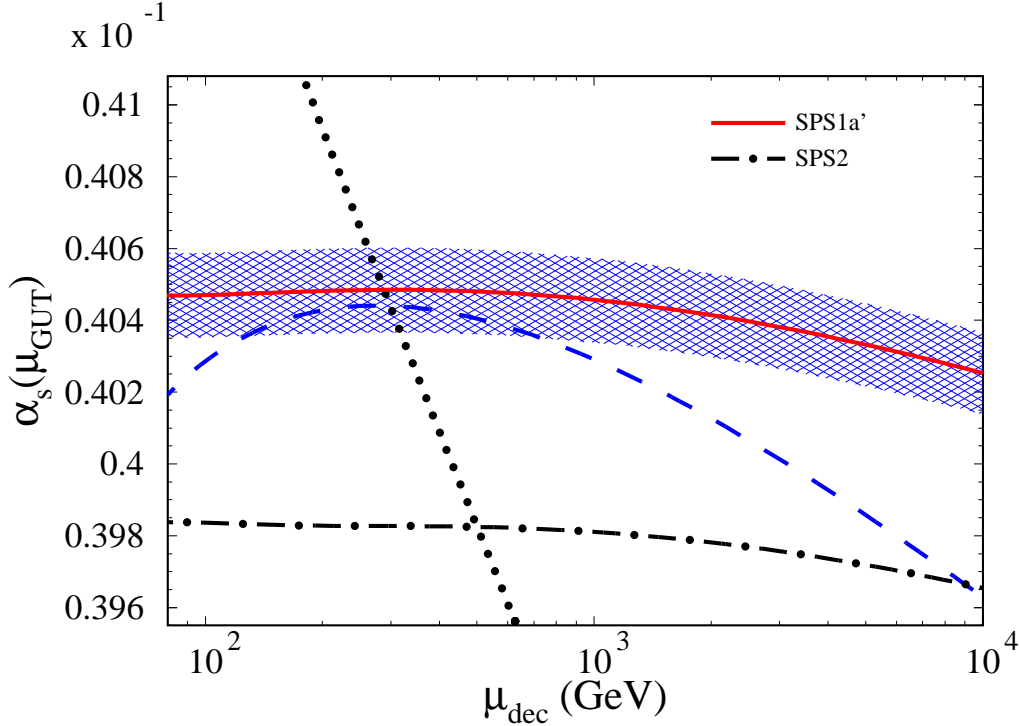,width=36em}
  \caption{\label{fig::asGUT}
    $\alpha_s(\mu_{\rm GUT})$ as a function of $\mu_{\rm dec}$.
    Dotted, dashed and solid lines denote  the  one-, two-, and three-loop
    contributions, respectively, obtained by using for the input parameters their
    values for the SPS1a$^\prime$ benchmark point. The dash-dotted line shows the
    three-loop running corresponding to the SPS2 point.
  }
\end{figure}

\begin{figure}[ht]
\epsfig{figure=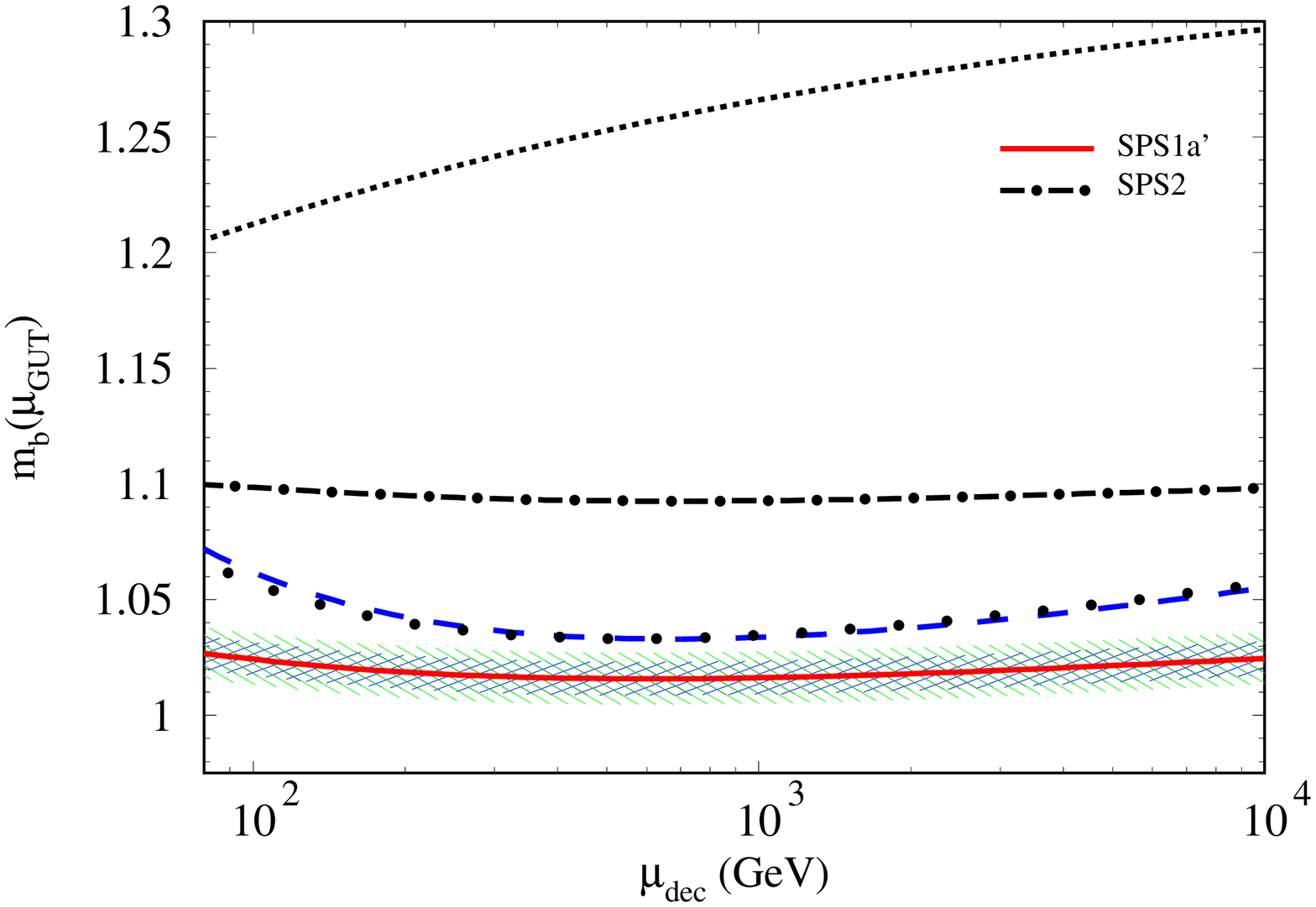,width=36em}
  \caption{\label{fig::mbGUT}
    $m_b(\mu_{\rm GUT})$ as a function of $\mu_{\rm dec}$.
   The  fine-dotted, dashed and solid lines denote  the  one-, two-, and three-loop
    contributions to the running bottom mass for the
    SPS1a$^\prime$ benchmark point, respectively.  The 
    dotted line displays the two-loop running, where the
    $\tan\beta$ enhanced 
    contributions are resummed according to
    Ref.~\cite{Carena:1999py}. The dash-dotted line represents the three-loop
    running corresponding to the SPS2 benchmark point.
  }
\end{figure}

The  decoupling scale is not a physical parameter and cannot be
predicted by the theory. It is usually chosen to be of the order of the  heavy
particle mass  in order to circumvent the appearance of large
logarithms.  At fixed order perturbation theory, it is expected that the
relations between the running parameters 
evaluated at high-energy scales and their low-energy values become less
sensitive to the choice of $\mu_{\rm dec}$ once higher order radiative
corrections are considered. The dependence on the precise value of the
decoupling scale 
is interpreted as a measure of the unknown higher order corrections. We discuss
the scale dependence of  $\alpha_s(\mu_{\rm GUT})$ and $m_b(\mu_{\rm GUT})$ 
in Fig.~\ref{fig::asGUT} and
Fig.~\ref{fig::mbGUT}, respectively.

For the SM parameters we  used 
$\alpha_s(M_Z)=0.1189$~\cite{Bethke:2006ac}, where $ M_Z =
91.1876~\mbox{GeV}\,$~\cite{Yao:2006px}, 
$m_b(\mu_b)=4.164$~GeV~\cite{Kuhn:2007vp}, with $\mu_b=m_b(\mu_b)$, and $M_t =
172.4~\mbox{GeV}\,$~\cite{:2008vn}. For the SQCD parameters, we
implemented  their values for the SPS1a$^\prime$
scenario~\cite{Aguilar-Saavedra:2005pw}: $\mg=607.1$~GeV,
$m_{\tilde{t}_1}=366.5$~GeV, $m_{\tilde{t}_2}=585.5$~GeV,
$m_{\tilde{b}_1}=506.3$~GeV, $m_{\tilde{b}_2}=545.7$~GeV, $A_t^{\drbarmath{}}
(1~{\mbox{TeV}})=-565.1$~GeV, $A_b^{\drbarmath{}}
(1~{\mbox{TeV}})=-943.4$~GeV, $\mu = 396.0$~GeV, and $\tan\beta=10.0$ . 

For the calculation of $\zeta_{m_b}$ to two-loop accuracy, the \drbar{} parameter 
 $A_b$ has to be converted to the renormalization scheme we used here.\footnote{See
  Ref.~\cite{Heinemeyer:2004xw} for a detailed discussion.} 
For the accuracy level we are considering, the one-loop conversion
relation is required
\begin{eqnarray}
A_b^{\rm mixed} = A_b^{\drbarmath{}}+ \Delta A_b\,, \quad
\mbox{where}\quad \Delta A_b =\delta A_b^{\drbarmath{}} -
\delta A_b^{\rm mixed}\,.
\end{eqnarray}
The counterterms $\delta A_b^{i}$ were defined in Eq.~(\ref{eq::dab})
and the superscript $i$ indicates the renormalization scheme. The
shift $\Delta A_b$ is a finite quantity as it can be explicitly checked.
It depends in turn on  the running bottom quark
mass in the MSSM.  We use an iterative method and choose the running
bottom quark 
mass in the SM as the initial parameter. A stable solution is
obtained  after few iterations. In addition, the energy evolution of the
parameter $A_b$ has to be taken into account. We use here the
one-loop RGE, that can be derived from Eq.~(\ref{eq::dab}).

The dependence on the decoupling scale for $\alpha_s(\mu_{\rm GUT})$ 
 is displayed in Fig.~\ref{fig::asGUT}. 
 The dotted, dashed and solid lines denote the one-, two-, and
 three-loop running, where the corresponding exact results for the
 decoupling coefficients have been implemented. One can see the improved
 stability of the three-loop results  w.r.t.\  the decoupling-scale
 variation. The uncertainty  induced by the current experimental
accuracy  on $\alpha_s(M_Z)$, $\delta \alpha_s=
0.001$\cite{Bethke:2006ac},  is indicated by the 
hatched band.\\
In order to get an idea of the effects induced by the SUSY mass parameters
on $\alpha_s(\mugut)$, we show through
 the dash-dotted line
  the three-loop results if  the SUSY parameters  corresponding to
 the Snowmass Point SPS2~\cite{Ghodbane:2002kg} are adopted. Their
 explicit values  are: $\mg=784.4$~GeV,
$m_{\tilde{t}_1}=1003.9$~GeV, $m_{\tilde{t}_2}=1307.4$~GeV,
$m_{\tilde{b}_1}=1296.6$~GeV, $m_{\tilde{b}_2}=1520.1$~GeV, and $\tan\beta=10.0$. The curves
induced by the other benchmark   points SPSi, with
$i=3,4,\ldots,9$ would lie between the two curves displayed here.
   One clearly notices  the great
impact of the SUSY-mass pattern on the predicted value of the strong
coupling at high energies.  Accordingly, for precision studies the explicit mass
pattern of heavy particles must  be taken into account.

In Fig.~\ref{fig::mbGUT} the scale dependence  for 
$m_b(\mu_{\rm GUT})$ is shown. The fine-dotted, dashed and solid lines
correspond to the exact one-, two-, and three-loop running obtained in
the SPS1a$^\prime$ scenario. 
As explained above, the energy
evolution of the running parameters have to be combined with the appropriate
matching conditions  between the low- and high-energy regimes. More
explicitly, in case of $m_b$ we determine its value within SQCD at the
energy-scale $\mu_{\rm dec}$ through the relation
\begin{eqnarray}
m_b^{\rm SQCD}(\mu_{\rm dec})= \frac{m_b^{(5)}(\mu_{\rm
    dec})}{\zeta_{m_b}(\mu_{\rm dec})}\,,\quad \mbox{where}\quad
\frac{1}{\zeta_{m_b}}=\frac{1}{1+\delta\zeta_{m_b}^{\tan\beta}+
  \delta\zeta_{m_b}^{\rm rest}}\,. 
\label{eq::resumm}
\end{eqnarray}
Here $\delta\zeta_{m_b}^{\tan\beta}$ denotes the contributions proportional with
$\tan\beta$ and  $\delta\zeta_{m_b}^{rest}$ the remaining
corrections. For simplicity, we do not show in Eq.~(\ref{eq::resumm}) the
explicit dependence on the MSSM parameters. For the $i$-loop running
analysis, we take into account  the  $(i-1)$-loop contribution to the
Eq.~(\ref{eq::resumm}). As can be
seen from the Figure~\ref{fig::mbGUT}, the three-loop 
results  stabilize the scale dependence and  reduce further the
theoretical uncertainty.  
\\
The dotted line displays the  two-loop running bottom-mass, where
the  contributions proportional with $\tan\beta$ to the one-loop $\zeta_{m_b}$ are
resummed following the method proposed in Refs.~\cite{Carena:1999py,
  Aguilar-Saavedra:2005pw}. Within this approach, the matching condition
can be written as 
\begin{eqnarray}
m_b^{\rm SQCD}(\mu_{\rm dec})= \frac{m_b^{(5)}(\mu_{\rm
    dec})}{\zeta_{m_b}^{\rm 1-loop}(\mu_{\rm dec})}\quad \mbox{and}\quad
\frac{1}{\zeta_{m_b}^{\rm 1-loop}} =
\frac{1- \delta\zeta_{m_b}^{\rm rest,
    1-loop}}{1+\delta\zeta_{m_b}^{\tan\beta,\rm{ 1-loop}}}\,.
\label{eq::resummc}
\end{eqnarray}
The superscript ${\rm 1-loop}$ indicates the order in perturbation
theory at which the individual contributions are evaluated.
The authors of Ref.~\cite{Carena:1999py} showed that, for a consistent
analysis not only the 
$\tan\beta$-enhanced contributions have to be resummed, but also
the next-to-leading logarithms (NLL)
$\alpha_s^{i+1}\ln^i(\mu^2/m_b^2)$. In our approach based
on $i$-loop RGEs and $(i-1)$-loop decoupling coefficients  the
 NLL are  implicitly resummed. 
The very good agreement between the two computations can be explained by the
fact that at one-loop order $\delta\zeta_{m_b}^{rest}$ is almost an
order of magnitude smaller than $\delta\zeta_{m_b}^{\tan\beta}$. 
 \\
 The experimental uncertainty generated by $\delta
\,\alpha_s=0.001$\cite{Bethke:2006ac} corresponds to the wider hatched
band, and 
the one due to  $\delta\, m_b = 25$~MeV\cite{Kuhn:2007vp} to the narrow
band. 
Let us notice that the three-loop order effects
exceed the uncertainty due to current experimental accuracy
$\delta\alpha_s$. 
\\
Finally, the dash-dotted line shows the three-loop running if the  SPS2 scenario
is implemented. The differences between  the three-loop order results
are mainly due to the change of masses of the SUSY particles. 

\begin{table}[h]
{\scalefont{1.0}
\begin{center}
\begin{tabular}{l lllll}
\hline
\hline
\multicolumn{6}{c}{$\mu_{\rm ren}=1000$~GeV} \\
\hline
$\alpha_s(\mu_{\rm ren})$
 &  0.0929 & $\pm 0.0006|_{\delta\,\alpha_s(M_Z)}$ & & $-
 0.003|_{\rm SPS2}$ &  $\pm 0.0001|_{\rm th}$\\ 
$m_b(\mu_{\rm ren})$
 &  2.164  &  $\pm 0.017|_{\delta\,\alpha_s(M_Z)}$ & $\pm
 0.015|_{\delta\, m_b(m_b)}$ &  $+ 0.12|_{\rm  SPS2}$ &   $\pm 0.01|_{\rm th}$ 
\\
\hline
\\
\hline
\hline
\multicolumn{6}{c}{$\mu_{\rm ren}=\mu_{\rm GUT}$} \\
\hline
$\alpha_s(\mu_{\rm ren})$
 &0.0405   & $\pm 0.0001|_{\delta\,\alpha_s(M_Z)}$ & & $\pm
 0.0007|_{\rm SPS2}$ &  $\pm 0.0001|_{\rm th}$\\ 
$m_b(\mu_{\rm ren})$
 & 1.016   &  $\pm 0.011|_{\delta\,\alpha_s(M_Z)}$ & $\pm
 0.007|_{\delta\, m_b(m_b)}$ &  $+ 0.077|_{\rm SPS2}$ &   $\pm 0.005|_{\rm th}$ 
\\
\hline
\end{tabular}
  \caption{\label{tab::res1}
    Numerical results for the strong coupling and bottom-quark  mass for
    $\mu_{\rm ren}=1000$~GeV and $\mu_{\rm ren}=\mu_{\rm GUT}$, respectively. The
    experimental inaccuracy 
    is evaluated taking into account the present uncertainty on $\alpha_s(M_Z)$ and $
    m_b(m_b)$. The effects of the SUSY-mass parameters are evaluated
    w.r.t.\ the SPS2 benchmark point. The
     theoretical uncertainties due to unknown higher order corrections
     are estimated from the variation with the decoupling scale for
     $100\,\mbox{GeV}\le\mu_{\rm dec}\le 1\,\mbox{TeV} $.
    }
\end{center}
}
\end{table}

For quantitative comparison, we give in Table~\ref{tab::res1} the numerical
values for $\alpha_s(\mu_{\rm ren})$ and $m_b(\mu_{\rm ren})$ for $\mu_{\rm ren}=1000$~GeV and $\mu_{\rm ren}=\mu_{\rm
  GUT}$, evaluated with three-loop accuracy.  For
 the decoupling scale we choose    $\mu_{\rm dec}= 600$~GeV as  at this
 scale the difference 
between the two- and three-loop order corrections
 reaches a minimum.
The different sources of
uncertainties are explicitly displayed. The theoretical uncertainties
due to unknown  higher order corrections are
estimated from the variation of the three-loop
results modifying  the decoupling scale  from $100$~GeV to
$1$~TeV. The effects of the SUSY-mass parameters are evaluated as the
difference between the three-loop results corresponding to the
benchmark points SPS1a$^{\prime}$ and SPS2.
One can easily see that the impact of the 
SUSY-mass pattern is at least five times larger than the experimental accuracy.  
\\

As already pointed out in the previous sections the unification of the
Yukawa couplings is very sensitive to the MSSM parameters.  The dependence
on the soft SUSY breaking parameters is induced in our approximation only through the
decoupling coefficients. They comprise an explicit dependence through the $X_b$
parameter (in the case of $\zeta_{m_b}$) and an implicit one through the squark masses.

The analytical
formulae for $\zeta_s$ and $\zeta_m$ given in Section~\ref{sec::twoloop}
are expressed in terms of the 
physical squark masses. Since they are not known experimentally, for the
following 
numerical analyses,  we computed them making the 
assumption that the soft SUSY breaking mass parameters defined in the
on-shell scheme  obey 
the following relation
$M_{\tilde Q}(t)=M_{\tilde{D}}(\tilde{M})=M_{\tilde{U}}
(\tilde{M})=A_f=\pm\mu=\tilde{M}$,  
where 
$M_{\tilde Q}(t)$ is the on-shell input parameter in the stop-mass matrix.\footnote{See
  Section~\ref{sec::ren}  for definitions and Refs.~\cite{Bartl:1997yd, Djouadi:1998sq} for a
  comprehensive discussion.} The
corresponding input parameter in the sbottom-mass matrix
acquires a finite shift of ${\cal  O}(\alpha_s)$ \cite{Bartl:1997yd}.
Upon diagonalization of the squark-mass matrices Eq.~(\ref{eq::mixing}),
one obtains  the 
on-shell squark masses $m_{\tilde{q}_{1,2}}$. The parameter $A_b$ entering
through $X_b$ the 
one-loop results have to be converted from the on-shell scheme in the
 renormalization scheme we introduced in Section~\ref{sec::frame}.

\begin{figure}
  \begin{center}
    \begin{tabular}{c}
      \epsfig{figure=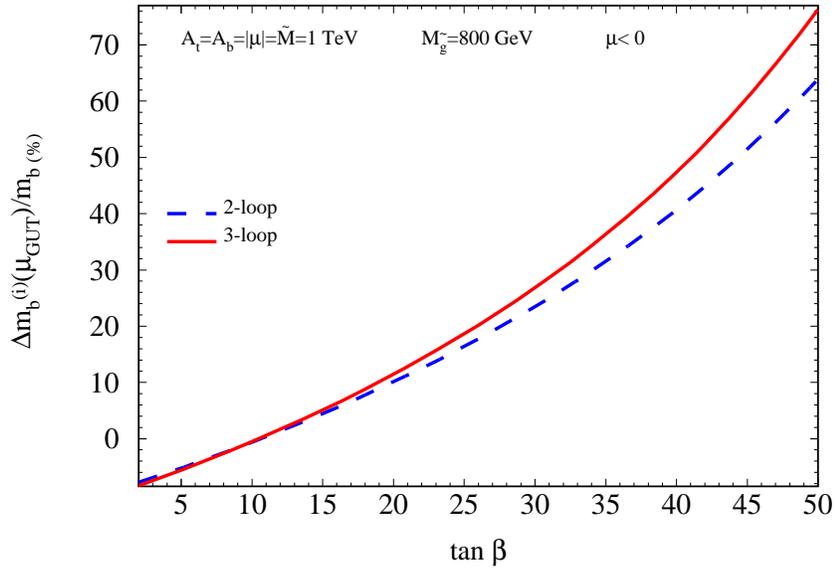,width=30em}
      \\(a)\\
      \epsfig{figure=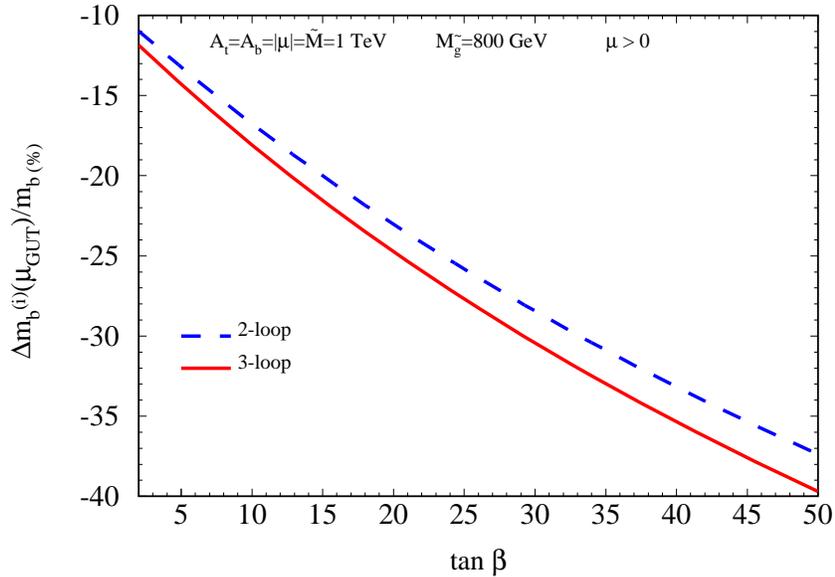,width=30em}
      \\  (b)
    \end{tabular}
    \parbox{14.cm}{
      \caption[]{\label{fig::tbeta}\sloppy
        $\Delta \,m_b/m_b$ as a function of $\tan\beta$ for $\mu<0$ (a)
        and  $\mu>0$ (b). The soft SUSY breaking mass parameters are fixed to $1$~TeV and
the gluino mass to $M_{\tilde g}=800$~GeV. 
        }}
  \end{center}
\end{figure}

\begin{figure}
  \begin{center}
    \begin{tabular}{c}
      \epsfig{figure=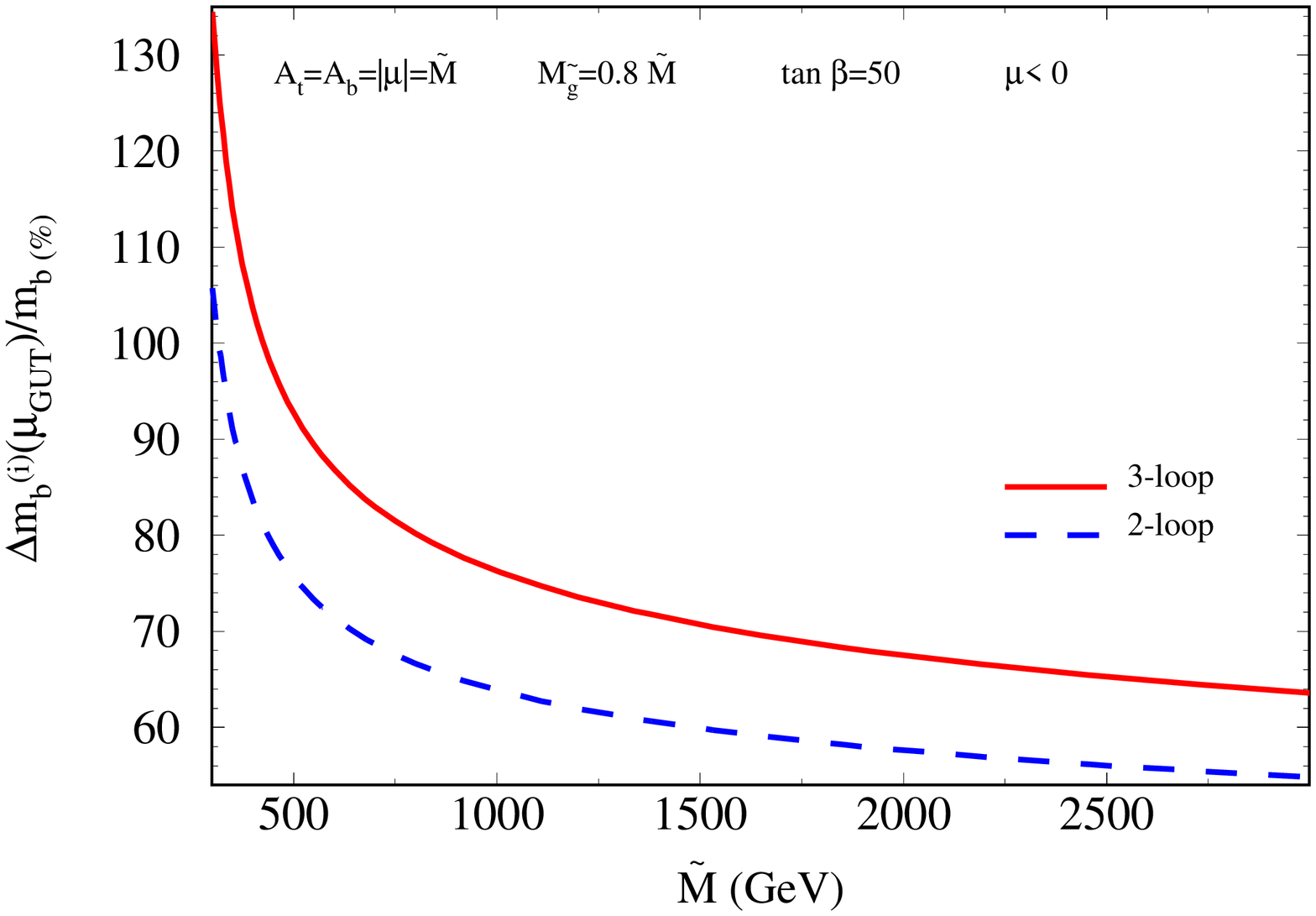,width=30em}
      \\(a)
      \epsfig{figure=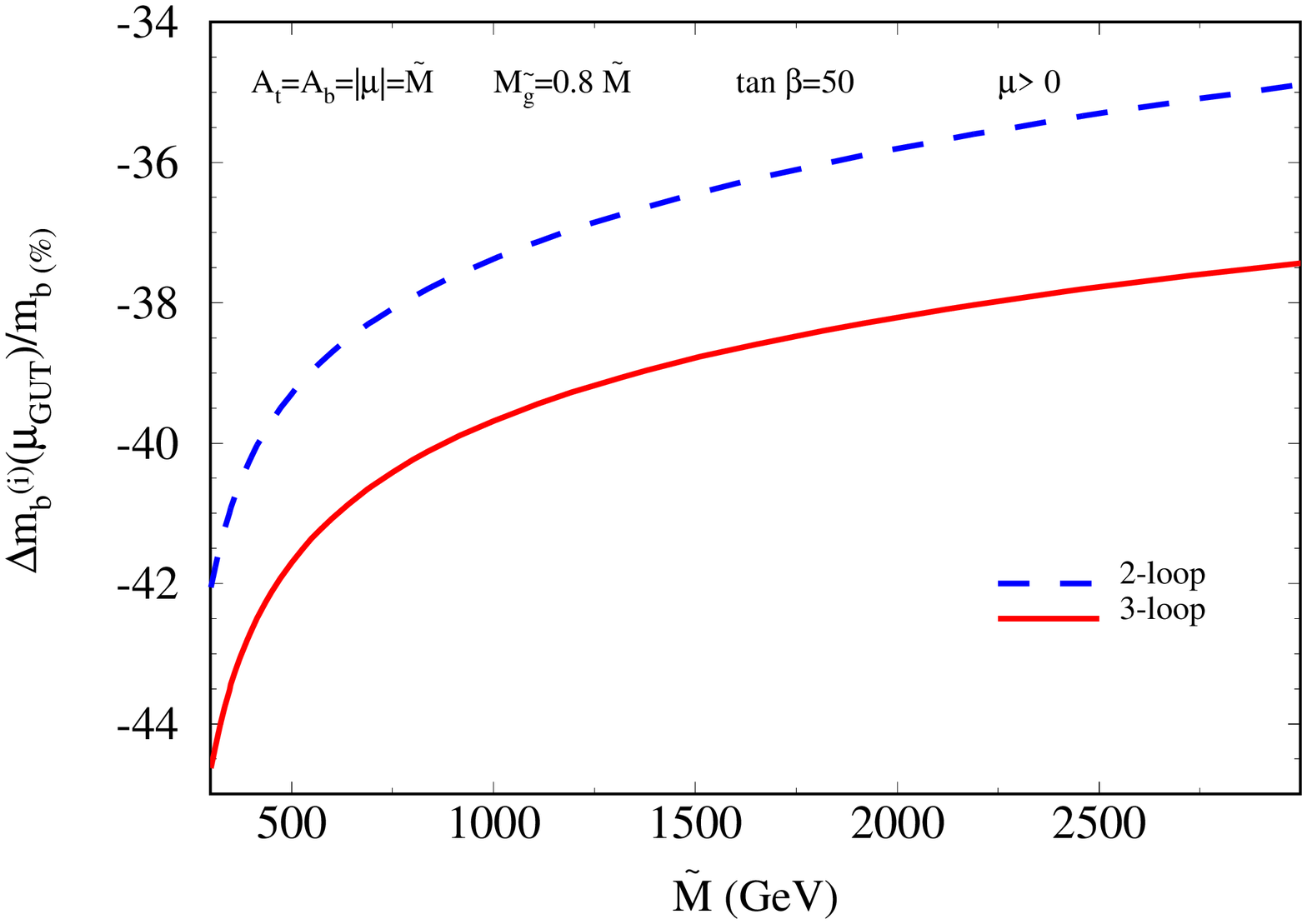,width=30em}
      \\(b)
    \end{tabular}
    \parbox{14.cm}{
      \caption[]{\label{fig::msusy}\sloppy
        $\Delta \,m_b/m_b$ as a function of $\tilde{M}$ for $\mu<0$ (a)
        and  $\mu>0$ (b). $\tan\beta$ is fixed to 50 and the gluino mass
        is $M_{\tilde g}=0.8 \tilde{M}$.
        }}
  \end{center}
\end{figure}  

In order to estimate the phenomenological impact, we discuss the
difference between 
 $m_b(\mu_{\rm GUT})$ evaluated using $i$-loop running and  $(i-1)$-loop
decoupling and the one-loop result
\begin{eqnarray}
\frac{\Delta \, m_b^{\mbox{(i)}}}{m_b}=\frac{m_b^{\mbox{(i-loop)}}  -
  m_b^{\mbox{(1-loop)}}}{m_b^{\mbox{(1-loop)}}}\,. 
\end{eqnarray}

In Fig.~\ref{fig::tbeta} we   fix the soft SUSY breaking mass
parameters to $\tilde{M}=1$~TeV and $M_{\tilde g}=800$~GeV, and study
the dependence of $\Delta \, 
m_b^{\mbox{(i)}}$ for $i=2$ (dashed) and $i=3$ (solid line) as a
function of  $\tan\beta$. One can clearly see the abrupt
increase/decrease  of the two- and 
three-loop order radiative corrections with the increase of
$\tan\beta$. Whereas the effects of the 
 the one-loop decoupling  can be as large as $65$\% for $\mu <0$
 and $-37$\% for   $\mu  >0$, the two-loop corrections are
 moderate, reaching at most $10$\% and $-3$\%, respectively.

The numerical
effects for large values of $\tan \beta$  are of special
interest for the study of Yukawa-coupling  unification. In
Fig.~\ref{fig::msusy} we show $\Delta \,m_b^{\mbox{(i)}}$ as a function
of the soft SUSY breaking mass scale for  $\tan\beta=50$. One can see the increase in
size of radiative corrections for lighter SUSY masses. Again, for $\mu
>0$ the bulk of the corrections are comprised in the two-loop running 
 mass, while the three-loop order effects sum up to few percent. For
 $\mu <0$ and light SUSY masses the three-loop contributions can increase the
 bottom-quark mass with up to $30$\%.

%- }}}

%- {{{conclusions
\section{\label{sec::concl}Conclusions}
The knowledge of  fundamental parameters at high energies, such
as $\tilde{M}$ or  $\mu_{\rm GUT}$, are essential for the reconstruction
of the theory beyond the SM. 
In this paper we presented the exact two-loop decoupling coefficients of the
strong coupling and the bottom-quark mass within the SQCD. Together with
the known three-loop
order RGEs they allow predictions of the two parameters at high
energies  with three-loop accuracy.  This level of precision on the theory side  is
necessary in order to match with the current experimental  accuracy. 
The   values of the gauge and Yukawa couplings at the unification scale
 $\mu_{\rm GUT}$ are essential ingredients for the determination of
the GUT threshold corrections, which in turn are used to identify the
underlying GUT model.\\ 
In addition, the dependence on the energy scale at which the supersymmetric particles
are integrated out, which reflects the size of the unknown higher order
corrections, is significantly reduced in the case of the three-loop
order predictions.  

Furthermore, the approach outlined here  accounts for the effects induced
by the individual mass parameters. They are phenomenologically
significant for both parameters and exceed the experimental
uncertainty by more than a factor five.
 
The radiative corrections to the running bottom-quark mass are
particularly important for SUSY models with large values of $\tan
\beta$. It turns out that for negative values of $\mu $ the
three-loop order contributions can reach up to $30\%$ from the tree-level 
bottom-quark mass. Furthermore, they clearly stabilize the perturbative
behaviour. These features render them  indispensable for studies
concerning the Yukawa-coupling unification. 

%- }}}
%- {{{ Acknowledgements:

\bigskip
\noindent
{\large\bf Acknowledgements}\\ 

We would like to thank M.~Steinhauser for his continuous support and
numerous  inspiring discussions and suggestions. We also thank him and R.~Harlander  for carefully
reading the manuscript. L.M. is grateful to
K.G.~Chetyrkin for  enlightening conversations concerning the decoupling
approach.  \\
We thank A.~V.~Bednyakov for providing us with the  results  necessary
for the numerical comparison with Ref.~\cite{Bednyakov:2007vm}. \\
This work was supported
by the DFG through SFB/TR~9 and by the Graduiertenkolleg
``Hochenergiephysik und Teilchenastrophysik''. 
%- }}}

%- {{{ bibliography:

%- }}}


\begin{thebibliography}{99}

%
% susy_dec.v2_ref.tex -- generated by sortref-2.3.6  
% ((C) R. Harlander, http://www.robert-harlander.de/software/)
% on Thu Oct 23 14:21:28 CEST 2008
%

%1
\bibitem{Dimopoulos:1981yj}
  S.~Dimopoulos, S.~Raby and F.~Wilczek,
  Phys.\ Rev.\  D {\bf 24}, 1681 (1981)

%2
\bibitem{Ibanez:1981yh}
  L.~E.~Ibanez and G.~G.~Ross,
  Phys.\ Lett.\ B {\bf 105}, 439 (1981)
  %%CITATION = PHLTA,B105,439;%%

%3
\bibitem{Amaldi:1991cn}
  U.~Amaldi, W.~de Boer and H.~F\"urstenau,
  Phys.\ Lett.\ B {\bf 260} (1991) 447
  %%CITATION = PHLTA,B260,447;%%

%4
\bibitem{Georgi:1974sy}
  H.~Georgi and S.~L.~Glashow,
  %``Unity Of All Elementary Particle Forces,''
  Phys.\ Rev.\ Lett.\  {\bf 32} (1974) 438.

%5
\bibitem{Fritzsch:1974nn}
  H.~Fritzsch and P.~Minkowski,
  Annals Phys.\  {\bf 93} (1975) 193

%5b
\bibitem{Langacker:1994bc}
  P.~Langacker and N.~Polonsky,
  Phys.\ Rev.\  D {\bf 50} (1994) 2199

%6
\bibitem{Hall:1993gn}
  L.~J.~Hall, R.~Rattazzi and U.~Sarid,
  Phys.\ Rev.\  D {\bf 50} (1994) 7048

%7
\bibitem{DiazCruz:2000mn}
  J.~L.~Diaz-Cruz, H.~Murayama and A.~Pierce,
  Phys.\ Rev.\  D {\bf 65}, 075011 (2002)

%8
\bibitem{Hempfling:1993kv}
  R.~Hempfling,
  Phys.\ Rev.\  D {\bf 49}, 6168 (1994)

%8b
\bibitem{Carena:1994bv}
  M.~S.~Carena, M.~Olechowski, S.~Pokorski and C.~E.~M.~Wagner,
  Nucl.\ Phys.\  B {\bf 426} (1994) 269

%9
\bibitem{Tobe:2003bc}
  K.~Tobe and J.~D.~Wells,
  Nucl.\ Phys.\  B {\bf 663}, 123 (2003)

%10
\bibitem{Blair:2002pg}
  G.~A.~Blair, W.~Porod and P.~M.~Zerwas,
  Eur.\ Phys.\ J.\ C {\bf 27} (2003) 263

%11
\bibitem{Aguilar-Saavedra:2005pw}
  J.~A.~Aguilar-Saavedra {\it et al.},
  Eur.\ Phys.\ J.\ C {\bf 46} (2006) 43

%12
\bibitem{Paige:2003mg}
  F.~E.~Paige, S.~D.~Protopopescu, H.~Baer and X.~Tata, [arXiv:hep-ph/0312045]

%13
\bibitem{Allanach:2001kg}
  B.~C.~Allanach,
  Comput.\ Phys.\ Commun.\  {\bf 143} (2002) 305

%14
\bibitem{Porod:2003um}
  W.~Porod,
  Comput.\ Phys.\ Commun.\  {\bf 153} (2003) 275

%15
\bibitem{Djouadi:2002ze}
  A.~Djouadi, J.~L.~Kneur and G.~Moultaka,
  Comput.\ Phys.\ Commun.\  {\bf 176} (2007) 426

%16
\bibitem{Martin:1993yx}
  S.~P.~Martin and M.~T.~Vaughn,
  Phys.\ Lett.\ B {\bf 318} (1993) 331 

%17
\bibitem{Martin:1993zk}
  S.~P.~Martin and M.~T.~Vaughn,
  Phys.\ Rev.\  D {\bf 50} (1994) 2282

%18
\bibitem{Jack:1994kd}
  I.~Jack and D.~R.~T.~Jones,
  Phys.\ Lett.\  B {\bf 333} (1994) 372

%19
\bibitem{Yamada:1994id}
  Y.~Yamada,
  Phys.\ Rev.\  D {\bf 50} (1994) 3537

%20
\bibitem{Pierce:1996zz}
  D.~M.~Pierce, J.~A.~Bagger, K.~T.~Matchev and R.~J.~Zhang,
  Nucl.\ Phys.\ B {\bf 491} (1997) 3 

%21
\bibitem{Allanach:2003jw}
  B.~C.~Allanach, S.~Kraml and W.~Porod,
  JHEP {\bf 0303} (2003) 016

%22
\bibitem{Harlander:2005wm}
  R.~Harlander, L.~Mihaila and M.~Steinhauser,
  Phys.\ Rev.\ D {\bf 72} (2005) 095009

%23
\bibitem{Harlander:2007wh}
  R.~V.~Harlander, L.~Mihaila and M.~Steinhauser,
  Phys.\ Rev.\  D {\bf 76} (2007) 055002

%24
\bibitem{Bethke:2006ac}
  S.~Bethke,
  Prog.\ Part.\ Nucl.\ Phys.\  {\bf 58} (2007) 351

%25
\bibitem{Kuhn:2007vp}
  J.~H.~K\"uhn, M.~Steinhauser and C.~Sturm,
  Nucl.\ Phys.\ B {\bf 778} (2007) 192

%26
\bibitem{Noth:2008tw}
  D.~Noth and M.~Spira,
  Phys.\ Rev.\ Lett.\  {\bf 101} (2008) 181801
%  arXiv:0808.0087 [hep-ph]

%27
\bibitem{Blazek:2002ta}
  T.~Blazek, R.~Dermisek and S.~Raby,
  Phys.\ Rev.\  D {\bf 65}, 115004 (2002)

%28
\bibitem{Carena:1999py}
  M.~Carena, D.~Garcia, U.~Nierste and C.~E.~M.~Wagner,
  Nucl.\ Phys.\ B {\bf 577} (2000) 88 

%29
\bibitem{Bardeen:1978yd}
  W.~A.~Bardeen, A.~J.~Buras, D.~W.~Duke and T.~Muta,
  Phys.\ Rev.\  D {\bf 18} (1978) 3998

%30
\bibitem{Siegel:1979wq}
  W.~Siegel,
  Phys.\ Lett.\  B {\bf 84} (1979) 193

%31
\bibitem{Appelquist:1974tg}
  T.~Appelquist and J.~Carazzone,
  Phys.\ Rev.\  D {\bf 11} (1975) 2856

%32
\bibitem{Chetyrkin:1997un}
  K.~G.~Chetyrkin, B.~A.~Kniehl and M.~Steinhauser,
  Nucl.\ Phys.\  B {\bf 510} (1998) 61

%33
\bibitem{Steinhauser:2002rq}
  M.~Steinhauser,
  Phys.\ Rept.\  {\bf 364} (2002) 247

%34
\bibitem{Ferreira:1996ug}
  P.~M.~Ferreira, I.~Jack and D.~R.~T.~Jones,
  Phys.\ Lett.\  B {\bf 387} (1996) 80

%35
\bibitem{Schroder:2005hy}
  Y.~Schroder and M.~Steinhauser,
  JHEP {\bf 0601}, 051 (2006)

%36
\bibitem{Chetyrkin:2005ia}
  K.~G.~Chetyrkin, J.~H.~Kuhn and C.~Sturm,
  Nucl.\ Phys.\  B {\bf 744} (2006) 121

%37
\bibitem{Bednyakov:2007vm}
  A.~V.~Bednyakov,
  Int.\ J.\ Mod.\ Phys.\  A {\bf 22} (2007) 5245

%38
\bibitem{Bednyakov:2002sf}
  A.~Bednyakov, A.~Onishchenko, V.~Velizhanin and O.~Veretin,
  Eur.\ Phys.\ J.\  C {\bf 29} (2003) 87

%39
\bibitem{Davydychev:1992mt}
  A.~I.~Davydychev and J.~B.~Tausk,
  Nucl.\ Phys.\  B {\bf 397}, 123 (1993)

%40
\bibitem{Chetyrkin:1981qh}
  K.~G.~Chetyrkin and F.~V.~Tkachov,
  Nucl.\ Phys.\  B {\bf 192}, 159 (1981)

%41
\bibitem{Nogueira:1991ex}
  P.~Nogueira,
  J.\ Comput.\ Phys.\  {\bf 105} (1993) 279

%42
\bibitem{Seidensticker:1999bb}
  T.~Seidensticker, hep-ph/9905298


%43
\bibitem{Vermaseren:2000nd}
  J.~A.~M.~Vermaseren,
  arXiv:math-ph/0010025

%44
\bibitem{Guasch:1998as}
  J.~Guasch, J.~Sola and W.~Hollik,
  Phys.\ Lett.\  B {\bf 437} (1998) 88

%45
\bibitem{Jack:1993ws}
  I.~Jack, D.~R.~T.~Jones and K.~L.~Roberts,
  Z.\ Phys.\  C {\bf 62} (1994) 161

%46
\bibitem{Jack:1994rk}
  I.~Jack, D.~R.~T.~Jones, S.~P.~Martin, M.~T.~Vaughn and Y.~Yamada,
  Phys.\ Rev.\  D {\bf 50} (1994) 5481

%47
\bibitem{Harlander:2006rj}
  R.~Harlander, P.~Kant, L.~Mihaila and M.~Steinhauser,
  JHEP {\bf 0609} (2006) 053

%48
\bibitem{Ellis:1975ap}
  J.~R.~Ellis, M.~K.~Gaillard and D.~V.~Nanopoulos,
  Nucl.\ Phys.\  B {\bf 106} (1976) 292;
\\
%\bibitem{Shifman:1978zn}
  M.~A.~Shifman, A.~I.~Vainshtein and V.~I.~Zakharov,
  Phys.\ Lett.\  B {\bf 78} (1978) 443;
\\
%\bibitem{Shifman:1979eb}
  M.~A.~Shifman, A.~I.~Vainshtein, M.~B.~Voloshin and V.~I.~Zakharov,
  Sov.\ J.\ Nucl.\ Phys.\  {\bf 30} (1979) 711
  [Yad.\ Fiz.\  {\bf 30} (1979) 1368];
\\
%\bibitem{Vainshtein:1980ea}
  A.~I.~Vainshtein, V.~I.~Zakharov and M.~A.~Shifman,
  Sov.\ Phys.\ Usp.\  {\bf 23} (1980) 429
  [Usp.\ Fiz.\ Nauk {\bf 131} (1980) 537];
\\
%\bibitem{Kniehl:1995tn}
  B.~A.~Kniehl and M.~Spira,
  Z.\ Phys.\  C {\bf 69} (1995) 77;
\\
%\bibitem{Kilian:1995tra}
  W.~Kilian,
  Z.\ Phys.\  C {\bf 69} (1995) 89;
\\
%\bibitem{Spira:1995rr}
  M.~Spira, A.~Djouadi, D.~Graudenz and P.~M.~Zerwas,
  Nucl.\ Phys.\  B {\bf 453} (1995) 17;

%49
\bibitem{Guasch:2003cv}
  J.~Guasch, P.~Hafliger and M.~Spira,
  Phys.\ Rev.\  D {\bf 68} (2003) 115001

%50
\bibitem{Yao:2006px}
  W.~M.~Yao {\it et al.}  [Particle Data Group],
  J.\ Phys.\ G {\bf 33} (2006) 1
 

%51
\bibitem{:2008vn}
    The Tevatron Electroweak Working Group and CDF Collaboration
                  and D0 Collaboration,
  arXiv:0808.1089 [hep-ex]

%52
\bibitem{Heinemeyer:2004xw}
  S.~Heinemeyer, W.~Hollik, H.~Rzehak and G.~Weiglein,
  Eur.\ Phys.\ J.\  C {\bf 39} (2005) 465

%53
\bibitem{Ghodbane:2002kg}
  N.~Ghodbane and H.~U.~Martyn,
  in {\it Proc. of the APS/DPF/DPB Summer Study on the Future of
    Particle Physics (Snowmass 2001) } ed. N.~Graf, 
  arXiv:hep-ph/0201233

%54
\bibitem{Bartl:1997yd}
  A.~Bartl, H.~Eberl, K.~Hidaka, T.~Kon, W.~Majerotto and Y.~Yamada,
  Phys.\ Lett.\  B {\bf 402} (1997) 303

%55
\bibitem{Djouadi:1998sq}
  A.~Djouadi, P.~Gambino, S.~Heinemeyer, W.~Hollik, C.~Junger and G.~Weiglein,
  Phys.\ Rev.\  D {\bf 57}, 4179 (1998)


\end{thebibliography}
\end{document}